\documentclass[11pt, a4paper]{article}

\pdfoutput=1
\usepackage{jheppub}
\usepackage{atlasphysics}
\usepackage{subfigure}
\usepackage{graphicx}% Include figure files

\usepackage{longtable}
\usepackage{multirow}
\usepackage{textcomp}
\usepackage{units}
\usepackage{marvosym}
\usepackage{hyperref}

\title{Measurement of  $W\gamma$ and $Z\gamma$ production in proton-proton collisions
       at $\sqrt{s}=$ 7 \TeV{} with the ATLAS Detector}% Force line breaks with \\

\author[1]{The ATLAS Collaboration \note{See Appendix for the list of collaboration members}}

\abstract{
We present studies of $W$ and $Z$ bosons with associated high energy photons
produced in $pp$ collisions at $\sqrt{s}$ = 7 TeV.  The analysis uses 35 pb$^{-1}$ of data
collected by the ATLAS experiment in 2010.
The event selection requires $W$ and $Z$ bosons decaying into high $p_{\mathrm{T}}$
leptons (electrons or muons) and a photon with $E_{\mathrm{T}}>15$ GeV separated from the lepton(s) 
by a distance $\Delta R(l,\gamma)>$ 0.7 in $\eta$-$\phi$ space.
A total of 95 (97)
$pp \to e^{\pm}\nu\gamma+X$ ($pp \to \mu^{\pm}\nu\gamma+X$) and 25 (23) 
$pp \to e^{+}e^{-}\gamma+X$ ($pp \to \mu^{+}\mu^{-}\gamma+X$)
event candidates are selected. The kinematic distributions of the leptons and photons and 
the production cross sections are measured. The data are found to
agree with Standard Model predictions that include next-to-leading-order $O(\alpha\alpha_s)$
contributions.
}


\begin{document}

\begin{flushright}
CERN-PH-EP-2011-079 \\
submitted to journal
\end{flushright}

\maketitle

%%%%%%%%%%%%%%%%%%%%%%%%%%%%%%%%%%%%
%            Content               %
%%%%%%%%%%%%%%%%%%%%%%%%%%%%%%%%%%%%

\clearpage

%%%%%%%%%%%%%%%%%%%%%%%%%%%%%%%%%%%%%%%%%%%%%%%%%%%
%% intro
%%%%%%%%%%%%%%%%%%%%%%%%%%%%%%%%%%%%%%%%%%%%%%%%%%%
%% \input{intro}        % 1.5 pages

\section{Introduction}
\label{sec:intro}

Measurements of the production of $W$ and $Z$ bosons with associated
high energy photons provide important tests of the Standard Model (SM)
of particle physics.  The $W\gamma$ process is directly sensitive to
the triple gauge boson couplings predicted by the non-Abelian $SU(2)_L
\times U(1)_Y$ gauge group of the electroweak sector.  The triple
gauge boson couplings in the $Z\gamma$ process vanish in the SM at
tree level.  Physics beyond the SM such as composite structure of $W $
and $Z $ bosons, new vector bosons, and techni-mesons would enhance
production cross sections and alter the event kinematics. Data taken
with the ATLAS detector~\cite{DetectorPaper:2008} provide a new
opportunity to study $W\gamma$ and $Z\gamma$ production using the high
energy $pp$ collisions provided by the Large Hadron Collider (LHC).
Previous hadroproduction measurements have been made at the Fermilab
Tevatron collider by the CDF~\cite{CDFpaper} and D0~\cite{D0paper}
collaborations using $p\bar{p}$ collisions at $\sqrt{s}=1.96$~TeV and
at LHC by the CMS~\cite{Chatrchyan:2011rr} collaboration.
 
Our studies use measurements of
$pp \to l^{\pm}\nu\gamma+X$ and $pp \to l^+l^-\gamma+X$ production at
$\sqrt{s}=$ 7 TeV  with an integrated luminosity of approximately 35 pb$^{-1}$.
Events are selected by requiring the presence of a $W$ or $Z$ boson 
candidate along with an associated isolated photon having a transverse energy
$E_{\mathrm{T}} > 15$~GeV and separated from the 
closest electron or muon $l$ by $\Delta R(l,\gamma)>$ 0.7
\footnote{
The nominal interaction point is defined as the origin of the coordinate system,
while the anti-clockwise beam direction defines the z-axis and the $x-y$ plane is
transverse to the beam direction. The positive x-axis is defined as pointing from
the interaction point to the centre of the LHC ring and the positive y-axis is
defined as pointing upwards. The azimuthal angle $\phi$ is measured around the
beam axis and the polar angle $\theta$ is the angle from the beam axis.
The pseudorapidity is defined as $\eta=-$ln $tan(\theta/2)$.
The distance $\Delta R$ in the $\eta-\phi$ space is defined as
$\Delta R = \sqrt{\Delta \eta^2 + \Delta \phi^2}$}.

\begin{figure}[htbp]
  \centering
  \subfigure[u-channel]{\includegraphics[width=0.40\columnwidth]{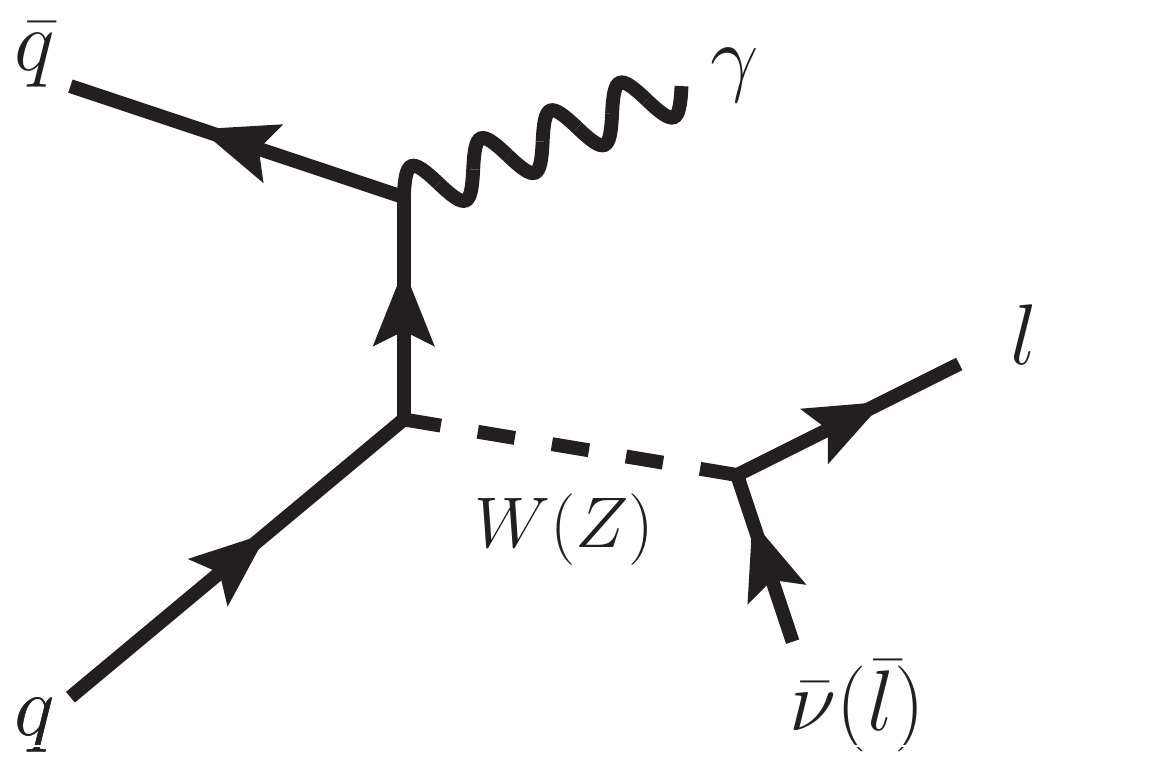}}
  \subfigure[t-channel]{\includegraphics[width=0.40\columnwidth]{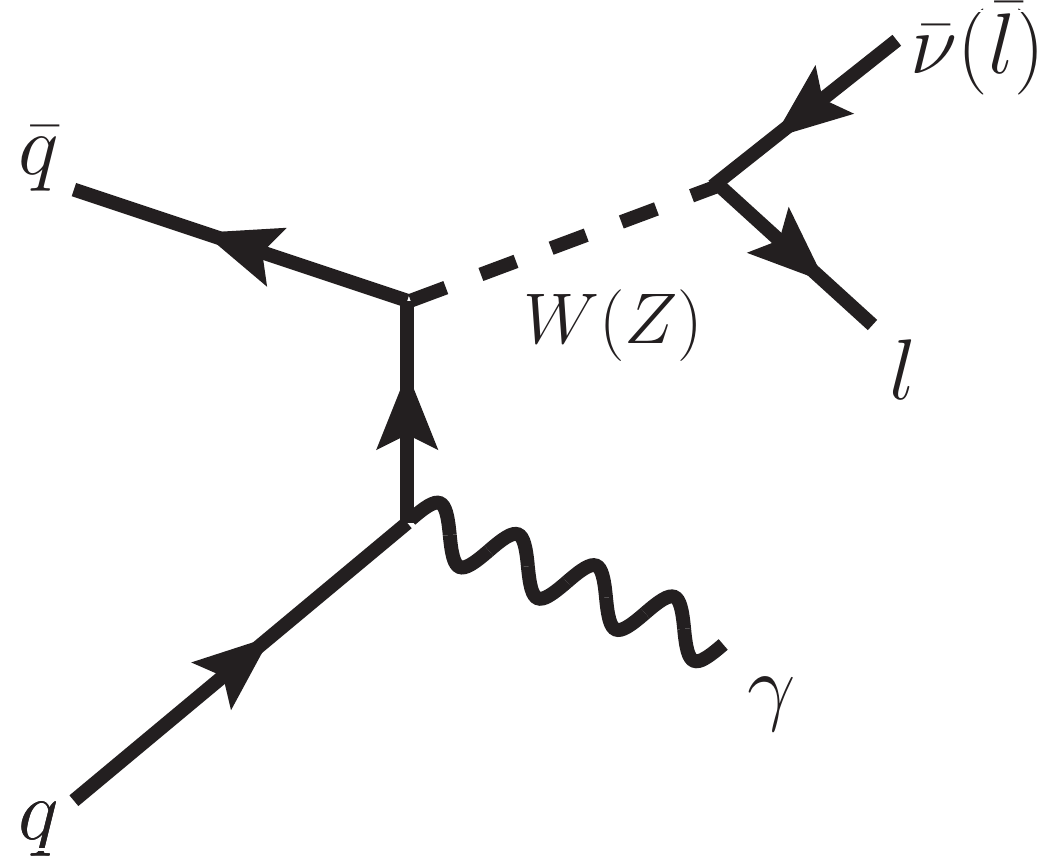}}
  \subfigure[FSR]{\includegraphics[width=0.40\columnwidth]{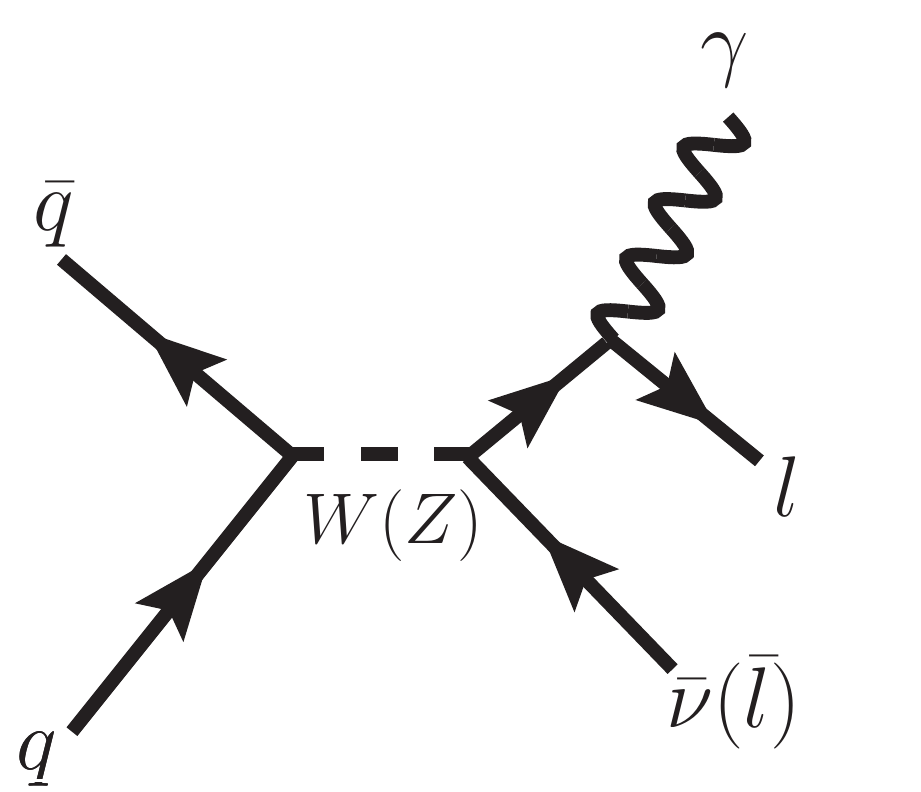}}
  \subfigure[s-channel]{\includegraphics[width=0.40\columnwidth]{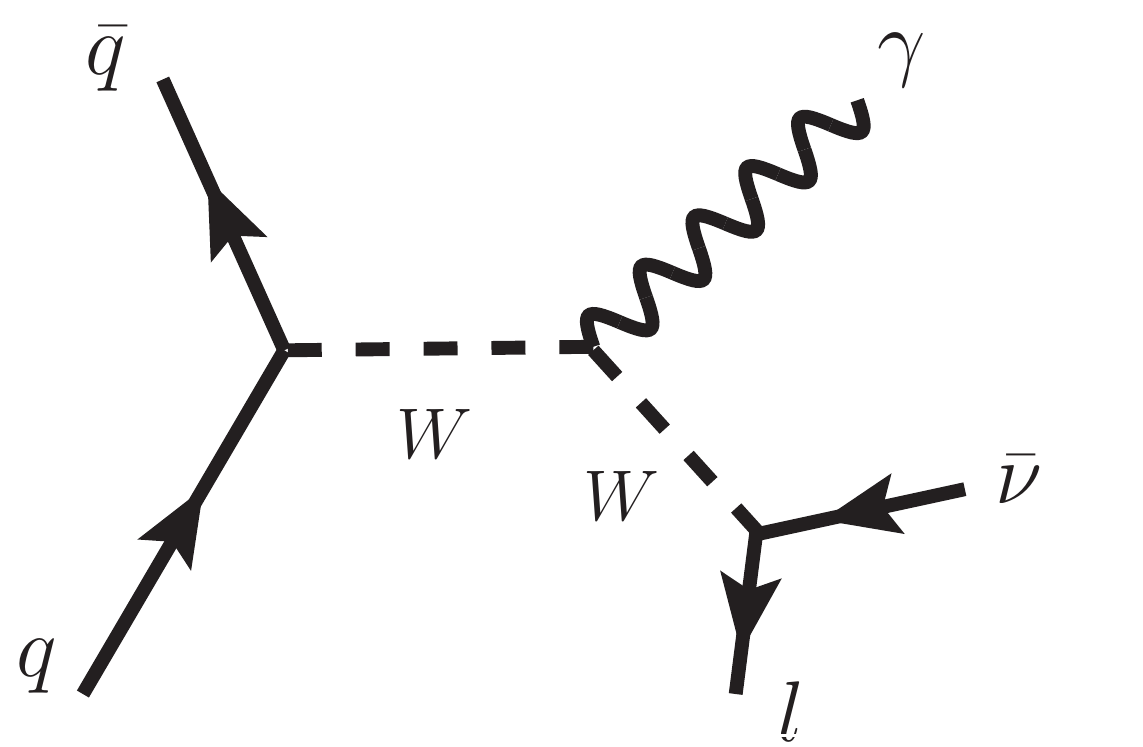}}
  \caption{Feynman diagrams of $W\gamma$ and $Z\gamma$ production in (a) u-channel
    (b) t-channel and (c) final state photon radiation (FSR) from the $W$ and $Z$ boson decay process.
    (d) Feynman diagram of $W\gamma$ production in the s-channel.}
  \label{fig:fey_dia}
\end{figure}

\begin{figure}[htbp]
  \centering
  \centering
  \subfigure[]{\includegraphics[width=0.40\columnwidth]{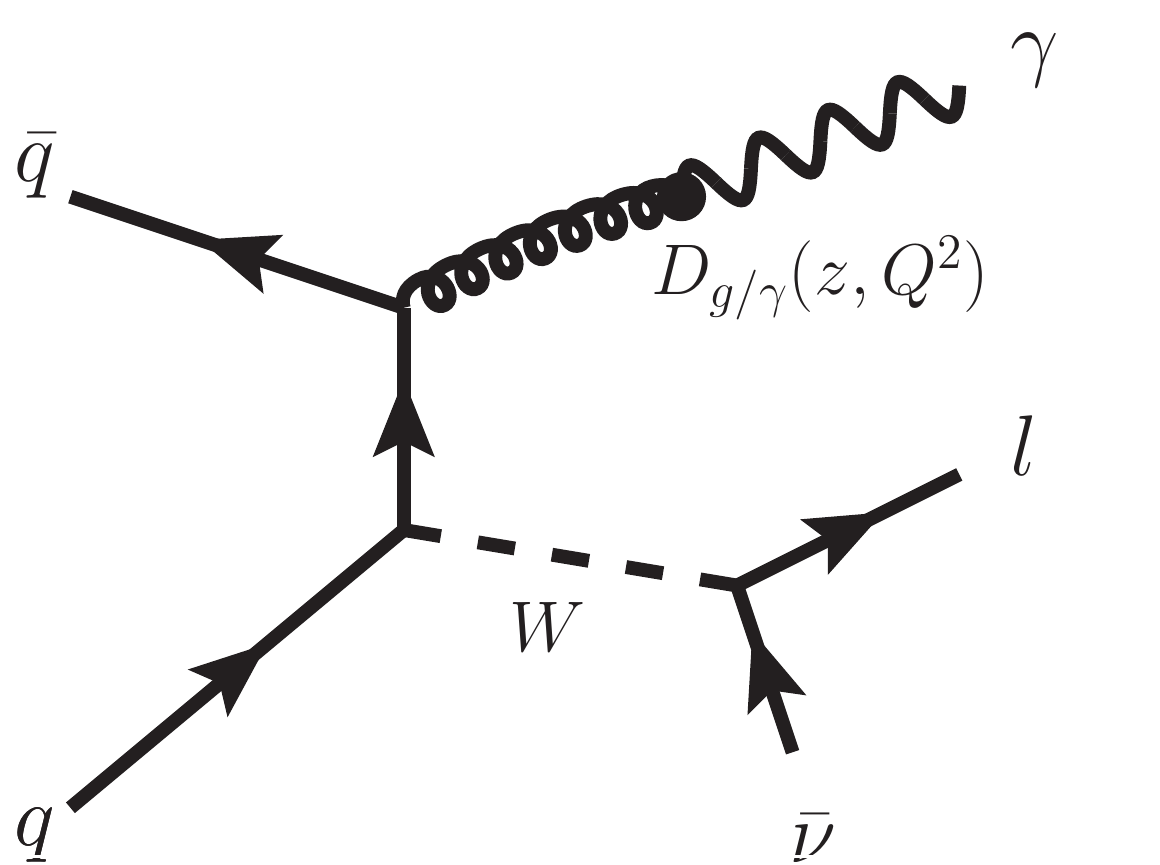}}
  \subfigure[]{\includegraphics[width=0.40\columnwidth]{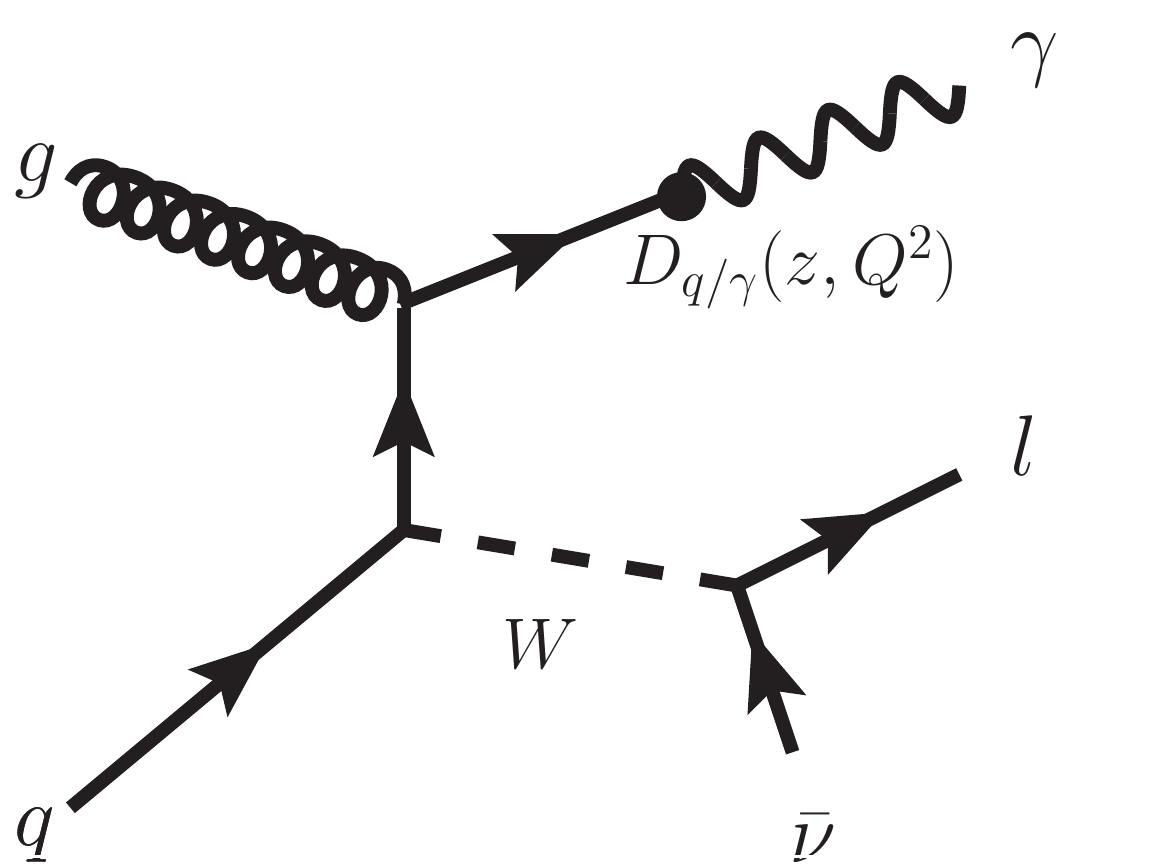}}
  \caption{Diagrams of the signal contributions from the
    $W+q(g)$~processes when a photon emerges from the fragmentation of the
    final state parton.}
  \label{fig:fragm_dia}
\end{figure}

The sources of the $l^{\pm}\nu\gamma$ and $l^{+}l^{-}\gamma$ final states are
$W\gamma \to l^{\pm}\nu\gamma$ and $Z\gamma \to l^+l^-\gamma$ production,
as well as QED final state radiation from inclusive $W$ and
$Z$ production:  $W \to l^{\pm}\nu \to l^{\pm}\nu\gamma$, $Z \to l^+l^- \to
l^+l^-\gamma$~(Fig.~\ref{fig:fey_dia}).
The data also include events with photons coming from hard fragmentation
of a quark or gluon~(see Fig.~\ref{fig:fragm_dia} for the case of $l\nu\gamma$). 
This source, while reduced by the photon identification and
isolation requirements, cannot be
neglected and is considered as a part of the signal process in the
analysis presented here.
Throughout this document the label ``$Z$'' refers to $Z/\gamma^*$
\footnote{$\gamma^{*}$ denotes an off-shell photon.}
and the notations $W\gamma$ and $Z\gamma$ are used to
denote the $l^{\pm}\nu\gamma$ and $l^+l^-\gamma$ final states.

%%%%%%%%%%%%%%%%%%%%%%%%%%%%%
% Figure
%%%%%%%%%%%%%%%%%%%%%%%%%%%%%

%%%%%%%%%%%%%%%%%%%%%%%%%%%%%
% Figure
%%%%%%%%%%%%%%%%%%%%%%%%%%%%%

\section{Monte Carlo Simulations of Standard Model Predictions for  the $W\gamma$ and
$Z\gamma$ Signal and Backgrounds}
\label{sec:theory_xsection}

%%%Our measurements are compared to SM predictions for $pp$ collisions
%%that  lead to the final states $l\nu\gamma + X$ and $ll\gamma + X$
%%where the photon is from direct $W\gamma$ and $Z\gamma$ di-boson production, from
%%final state radiation off the leptons in the $W/Z$ decays and from
%%quark/gluon fragmentation into an isolated photon. 
Monte Carlo (MC) event samples with full ATLAS detector simulation are used
for comparisons of the data to the theoretical expectations for the
$W\gamma$ and $Z\gamma$ signals and various backgrounds.
In this section the details of the MC event generators are described.

Since next-to-leading-order (NLO) generators with parton shower simulation are not available for
the $W\gamma$ and $Z\gamma$ signal processes,
they are generated with a {\sc madgraph}~\cite{madgraph} leading-order (LO) matrix-element 
generator interfaced to {\sc pythia}~\cite{pythia}
for gluon radiation and hadronization, and {\sc photos}~\cite{photos}
for photon radiation off the electron or muon in the W and Z decay.
The simulations of the signal processes using the {\sc madgraph} generator 
include interference effects between amplitudes, and effects from boson decay widths.
The matrix-element calculation uses the
leading-order parton distribution function (PDF) sets CTEQ6L1 \cite{CTEQ6l1},
and the corresponding ATLAS MC tune 2009~\cite{atlas_tune}.
Both the $W\gamma$ and $Z\gamma$ {\sc madgraph} samples are generated with photon
$E_{\mathrm{T}}>$ 10 GeV and $\Delta R(l,\gamma)>$ 0.5.

Fig.~\ref{fig:fey_dia} illustrates the dominant sources of
$W\gamma$ and $Z\gamma$ events.
The final state radiation (FSR) from $W\gamma$ ($Z\gamma$) events
are identified with a cut on the invariant mass of the lepton-neutrino
(opposite charged di-lepton) at the parton generator level.
Those $W\gamma$ ($Z\gamma$) events with $m(l\nu)< 74$ GeV ($m(ll)<85$ GeV)
are categorized as FSR.
The remaining events are identified as initial state radiation events (ISR).
The $W\gamma$ and $Z\gamma$ ISR events include those with photon radiation
from initial state quarks, and  for $W\gamma$ production, from the
$WW\gamma$ vertex(see Fig.~\ref{fig:fey_dia} (d)).
The division of the generated LO events into FSR and ISR categories 
is needed in order to apply the higher order perturbative corrections 
described below.

There are significant modifications to the LO electroweak $W\gamma$
and $Z\gamma$ cross sections due to QCD corrections, as in the case
of inclusive $W$ and $Z$ boson production. 
To introduce QCD corrections, our approach is to weight the fully simulated
LO MC events with NLO $k$-factors.
NLO predictions considering both QED and QCD vertices ($O(\alpha \alpha_S)$)
are determined using the Baur program~\cite{PhysRevD.48.5140,NLOpaper},
a matrix element parton generator with complete
next-to-leading-logarithm diagrams for $W\gamma$ and $Z\gamma$
production using narrow width approximations for the $W$ and $Z$ bosons.
The NLO Baur calculations for $W\gamma$ and $Z\gamma$ di-boson
production do not include FSR off the decay leptons.
Therefore a $k$-factor $k_{\mathrm{ISR}}$ determined by comparing the Born level and
the NLO Baur MC calculations, is applied to LO events identified as ISR 
as described above. For the FSR LO event weighting a $k_{\mathrm{FSR}}$ is determined using
an inclusive $W$/$Z$ NLO calculation with the
assumption that inclusively produced bosons have the same production dynamics as
those with radiation off the decay leptons.
To suppress photon signal contributions from quark/gluon fragmentation~\cite{fregphoton}
(see Fig.~\ref{fig:fragm_dia} for the case of
$l^{\pm}\nu\gamma$) isolation cuts are  applied to the photons selected in the
$W\gamma$ and $Z\gamma$ data and those from simulated quark/gluon fragmentation
in the NLO generator.
The events used for the NLO $k$-factor calculation and for the theoretical
cross section predictions are generated with $\epsilon_h< 0.5$,
where $\epsilon_h$ is an isolation criterion at generation level.
The variable $\epsilon_h$ ($\epsilon_h^p$) is used for the definition
of isolated photons, at the parton (particle)  level and is defined as
the ratio of the sum of energies carried by the partons (particles)
emerging from the quark/gluon fragmentation processes (excluding the photon)
to the energy carried by the fragmented photon.
The isolation criteria are applied using an $\eta-\phi$ cone of 0.4 centered
on the photon.
With these isolation cuts the quark/gluon fragmentation photons are estimated
to contribute about 8$\%$ of the photons in the generated $W\gamma$ and $Z\gamma$ events.

In comparing the data to SM signal predictions, the background processes considered are $W/Z$+jets,
$W\to \tau \nu$, $Z \to ll$ (background for the $W\gamma$), and $t\bar{t}$.
The backgrounds from the production of single-top, direct single photon,
dibosons ($WW/WZ/ZZ$) and QCD multi-jets are found to be negligible.
We use the {\sc powheg}~\cite{powheg} generator to simulate the
$t\bar{t}$ production, with {\sc pythia} used to model parton showers.
All other background sources are simulated with {\sc pythia}.
For comparison to data, the cross sections for the background processes are
normalized to the results of higher order QCD calculations.  All
signal and background samples were generated at $\sqrt{s}=7$ TeV, and then
processed with a {\sc geant4} simulation of the detector~\cite{atlassimu}. The MC samples
are simulated with on average two primary interactions but matched
to data-taking conditions by weighting each event to obtain the
primary vertex multiplicity distribution observed in data.

%%%%%%%%%%%%%%%%%%%%%%%%%%%%%%%%%%%%%%%%%%%%%%%%%%%
%% detector
%%%%%%%%%%%%%%%%%%%%%%%%%%%%%%%%%%%%%%%%%%%%%%%%%%%
%% \input{detector}

\section{The ATLAS Detector}
\label{sec:atlasdet}

The ATLAS detector~\cite{DetectorPaper:2008} consists of an
inner tracking system (inner detector, or ID) surrounded by a thin
superconducting solenoid providing a 2~T axial magnetic field,
electromagnetic (EM) and hadronic calorimeters and by a muon spectrometer
(MS).  The ID is composed of three subsystems. The pixel (closest to
the beam axis and with the highest granularity) and the silicon
microstrip (SCT) detectors cover the pseudorapidity range $|\eta|<2.5$, while
the Transition Radiation Tracker (TRT) has an acceptance range of
$|\eta|<2.0$.  The TRT provides identification information for
electrons (and as a consequence also for photons that convert to
electron-positron pairs) by the detection of transition radiation.
The electromagnetic calorimeter is a lead liquid-argon (LAr) detector
that is divided into one barrel ($|\eta|<1.475$) and two end-cap
components ($1.375<|\eta|<3.2$).  The calorimeter consists of three
longitudinal layers with the first (strip) having the highest
granularity in the $\eta$ direction, and the second collecting most
of the electromagnetic shower energy.  A thin presampler layer
covering the range $|\eta|<1.8$ is used to correct for the energy
lost by EM particles upstream of the calorimeter.
The transition region between the calorimeter and end-cap
($1.37 < |\eta| < 1.52$) is omitted for the detection of electrons and
photons in this analysis.  The hadronic calorimeter system, which
surrounds the electromagnetic calorimeter, is based on two different
detector technologies, with scintillator tiles or LAr as the active
media, and with either steel, copper, or tungsten as the absorber
material.
The MS is based on three large superconducting aircore toroid magnets,
a system of three stations of chambers for precise tracking
measurements in the range $|\eta|<2.7$, and a muon trigger system which
extends to the range $|\eta|<$ 2.4.

The ATLAS detector has a three-level trigger system.
The first level trigger is largely based on custom built
electronics that examine a subset of the total detector
information to decide whether or not to record each event,
reducing the data rate to below the design value of
approximately 75 kHz.
The subsequent two trigger levels run on a processor farm
and look at more detector information with greater precision.
They provide the reduction to a final data-taking rate designed to
be approximately 200 Hz.

%%%%%%%%%%%%%%%%%%%%%%%%%%%%%%%%%%%%%%%%%%%%%%%%%%%
%% Event Selection
%%%%%%%%%%%%%%%%%%%%%%%%%%%%%%%%%%%%%%%%%%%%%%%%%%%
%% \input{evesel}       % 1 pages

\section{Data Samples}
\label{sec:dataset}

Events in this analysis were selected by triggers requiring at least one
identified electron or muon candidate.
The electron and muon trigger configurations changed during the data
taking period in order to keep up with the increasing instantaneous
luminosity delivered by the LHC.  
The strictest trigger
selection criteria were applied in the last data taking period
where leptons reconstructed at the third level of the trigger system were
required to have $E_{\mathrm{T}}>$ 15 GeV (electrons) and $p_{\mathrm{T}}>$ 13 GeV (muons).
 Application of beam,
detector, and data-quality requirements resulted in a total integrated
luminosity of 35.1 pb$^{-1}$ (33.9 pb$^{-1}$) for the events collected
with the electron (muon) trigger.  The uncertainty on the absolute
luminosity determination is 3.4 \%~\cite{atlaslumi,atlaslumi2}.

\section{Reconstruction and Selection of $W\gamma$ and $Z\gamma$ Candidates}
\label{sec:evesel}

In this analysis the $W\gamma$ final state consists of an isolated electron or muon,
large missing transverse energy due to the undetected neutrino, and an isolated photon.
The $Z\gamma$ final state contains one pair of $e^{+}$$e^{-}$ or
$\mu^{+}$$\mu^{-}$ leptons and an isolated photon.  Collision events
are selected by requiring at least one reconstructed primary vertex
consistent with the average beam spot position and with at least three
associated tracks.
The efficiency to reconstruct the primary vertex for $W\gamma$ and $Z\gamma$
events is 100\%.
The selection criteria for electrons, muons and transverse energy follow
closely those used for the $W$ and $Z$ boson inclusive cross section analysis~\cite{WZpaper}.
The selection criteria for the photon are similar to those used for the analysis of
inclusive photon production~\cite{photonpaper}.

\subsection{Reconstruction of Electrons, Muons, Photons and Missing  Transverse Energy}
\label{sec:reco}

The muon candidates are reconstructed by associating the muon tracks in
the MS to the tracks in the ID~\cite{WZpaper}.
The combined track parameters of the muon candidates are derived using
a statistical approach based on their respective errors.
The selected muon candidate is a combined track from the primary vertex
with $p_{\mathrm{T}}>20$~\GeV{} and $|\eta|<2.4$,
and is isolated by requiring that the summed $p_{\mathrm{T}}$ of the tracks in a 0.4 radian cone
around the muon candidate is less than 20\% of the muon $p_{\mathrm{T}}$.
The $p_{\mathrm{T}}$ measured by the MS alone must be greater than 10~GeV.
A quality cut based on the difference in the $p_{\mathrm{T}}$ measured
independently in the ID and MS
is applied to improve the purity of the muon candidates.
To ensure a high quality track of the combined muon candidate,
a minimum number of hits in the ID is required~\cite{ATLAS_Wjet}.
For the $W\gamma$ measurement in the muon channel, at least one muon candidate
is required in the event, whereas for the $Z\gamma$ measurement, the
selected events must have exactly two oppositely charged muon candidates.

The electron candidates are reconstructed from an electromagnetic
calorimeter cluster associated with a reconstructed charged particle in
the ID. 
The electron identification algorithm,
which only considers electron candidates in the range $|\eta|<2.47$ and
excluding the region $1.37<|\eta|<1.52$,
combines calorimeter and tracking information and
provides three reference sets of selections
(``loose'', ``medium'' and ``tight'') with progressively stricter
identification criteria and stronger jet rejection~\cite{WZpaper}.
For the ``medium'' selection, information about the shower shape and width
of the cluster, the quality of the associated track,
and the cluster/track matching, as well as the energy deposited in the
hadronic calorimeter are used for the identification. The ``tight''
selection uses in addition the ratio of cluster energy to track momentum, the
particle identification potential of the TRT and stricter track
quality requirements to further reject charged hadrons and electrons
from photon conversions~\cite{WZpaper}.
A set of cuts on these discriminating
variables are identified to maximize the background rejection while
keeping a high electron signal efficiency. Such cuts are determined for different
pseudorapidity and $E_{\mathrm{T}}$ regions to maintain a high electron
efficiency across the detector and over the electron transverse energy range. 
The selection of $Z\gamma$ events requires two oppositely charged
``medium'' electrons with $E_{\mathrm{T}}>$ 20~\GeV{}.
For the $W\gamma$ selection one ``tight'' electron is required in the
event with $E_{\mathrm{T}}>$ 20 \GeV{}. 
The event is rejected if there is an additional ``medium'' electron candidate
present that passes the same kinematic cuts.

The photon candidates use clustered energy deposits in the EM
calorimeter in the range $|\eta|<2.37$ (excluding the region $1.37<|\eta|<1.52$)
and with $E_{\mathrm{T}}>15$~\GeV{}.
As for electrons, the photon identification is based on
discriminating variables computed from calorimeter information which
provides a good separation of signal from background. 
In particular the high granularity of the first (strip)
layer in the $\eta$ direction that covers up to $|\eta|<2.4$,
provides a very effective discrimination
between single photon and multiple-photon showers produced in
meson (e.g. $\pi^{0}$, $\eta$) decays.
A set of cuts on these discriminating variables is identified for different
pseudorapidity regions.
The cuts are applied separately for converted and unconverted photons to account for
the wider shower shapes of the former due to the opposite bending of the two legs
from the conversion in the solenoid magnetic field.
To further reduce the background due to photons from $\pi^{0}$ and $\eta$ decays,
an isolation requirement of $E_{\mathrm{T}}^{\mathrm{iso}}<5$ \GeV{} is applied.
$E_{\mathrm{T}}^{\mathrm{iso}}$ is the total transverse energy recorded in the
calorimeter (of both electromagnetic and hadronic systems)
in a cone of radius $\Delta R=0.4$ around the photon direction
(excluding a small window of $0.125 \times 0.175$ in the $\eta-\phi$
space which contains the photon energy deposit).
$E_{\mathrm{T}}^{\mathrm{iso}}$ is corrected for the leakage of the photon energy into
the isolation cone and the contributions from the underlying and
pile-up activities in the event~\cite{photonpaper}.

The reconstruction of the missing transverse energy ($E_{\mathrm{T}}^{\mathrm{miss}}$)
follows the definition in Ref.~\cite{WZpaper}.
The $E_{\mathrm{T}}^{\mathrm{miss}}$ calculation is based on the energy deposits
of calorimeter cells inside three-dimensional clusters. Corrections for hadronic to
electromagnetic energy scale, dead material, out-of-cluster energy as
well as muon momentum for the muon channel are applied.
Events that have sporadic calorimeter noise and non-collision backgrounds,
which can affect the $E_{\mathrm{T}}^{\mathrm{miss}}$ reconstruction,
are removed~\cite{jetclean}.

\subsection{Event Selection}
\label{sec:sele}

In addition to the presence of one high $p_{\mathrm{T}}$ lepton and one high $E_{\mathrm{T}}$
isolated photon, $W\gamma$
candidates are required to have $E_{\mathrm{T}}^{\mathrm{miss}}>25$ \GeV{} and the
transverse mass of the lepton-$E_{\mathrm{T}}^{\mathrm{miss}}$ system
$m_{\mathrm{T}}(l,\nu)>40$ \GeV{},
where $m_{\mathrm{T}}(l,\nu)= \sqrt{2p_{\mathrm{T}}(l) \cdot E_{\mathrm{T}}^{\mathrm{miss}} \cdot (1-\cos{\Delta{\phi}})}$,
and $\Delta{\phi}$ is the azimuthal separation between the directions of the lepton
and the missing transverse energy vector.
For $Z\gamma$ candidates, the invariant mass of the two opposite charged
leptons ($m_{l^{+}l^{-}}$) is required to be greater than 40 \GeV{}.
In both $W\gamma$ and $Z\gamma$ analyses, a $\Delta R(l,\gamma)>$ 0.7
cut is applied to suppress the contributions from FSR photons in the
$W$ and $Z$ boson decays.
A total of 192 $W\gamma$ candidates (95 in the
electron and 97 in the muon channel) and 48 $Z\gamma$
candidates (25 in the electron and 23 in the muon channel) pass all
the requirements.

\subsection{Kinematic Distributions of Event Candidates}
\begin{figure*}
  \centering
  \subfigure[]{\includegraphics[width=0.49\columnwidth]{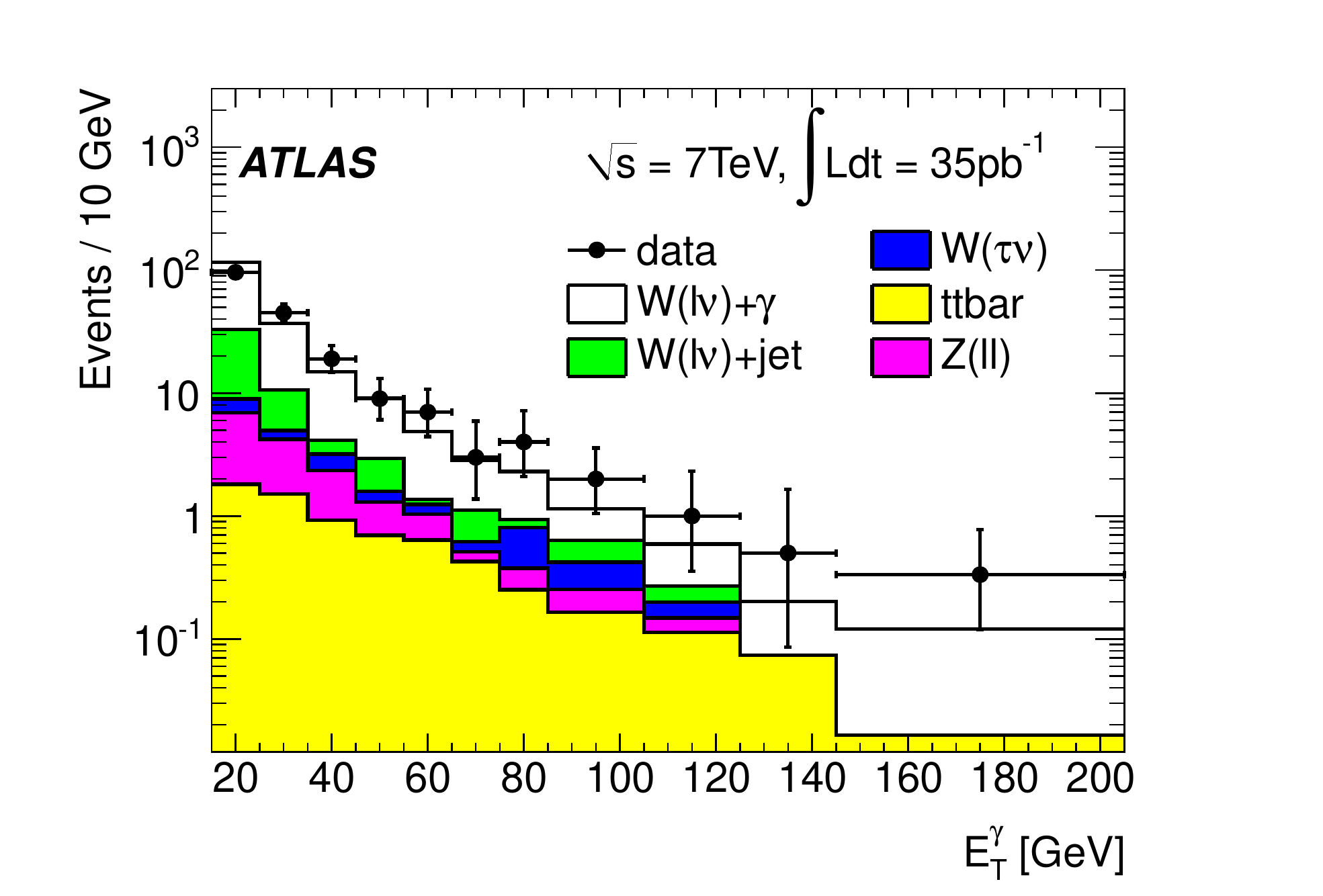}}
  \subfigure[]{\includegraphics[width=0.49\columnwidth]{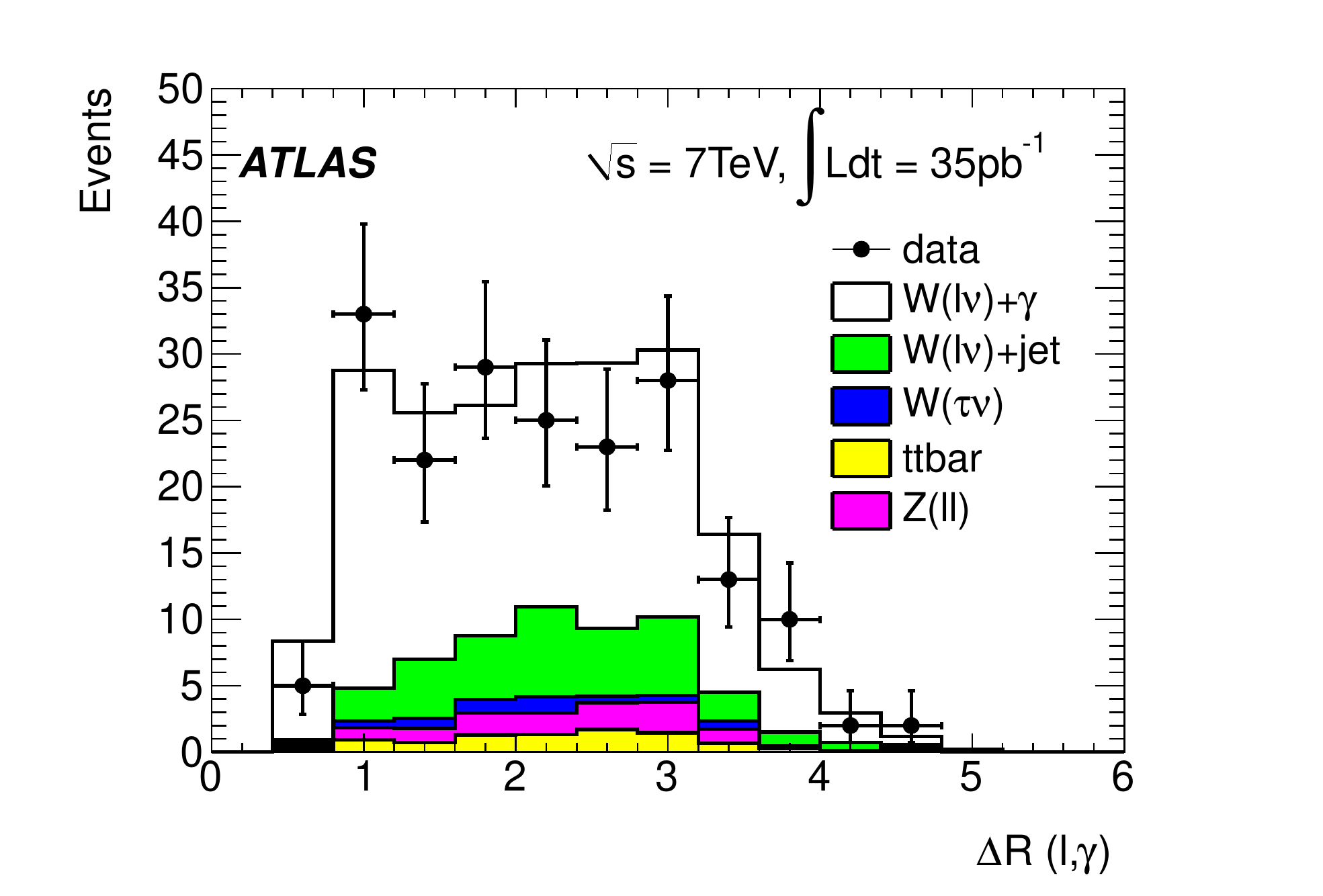}}
  \subfigure[]{\includegraphics[width=0.49\columnwidth]{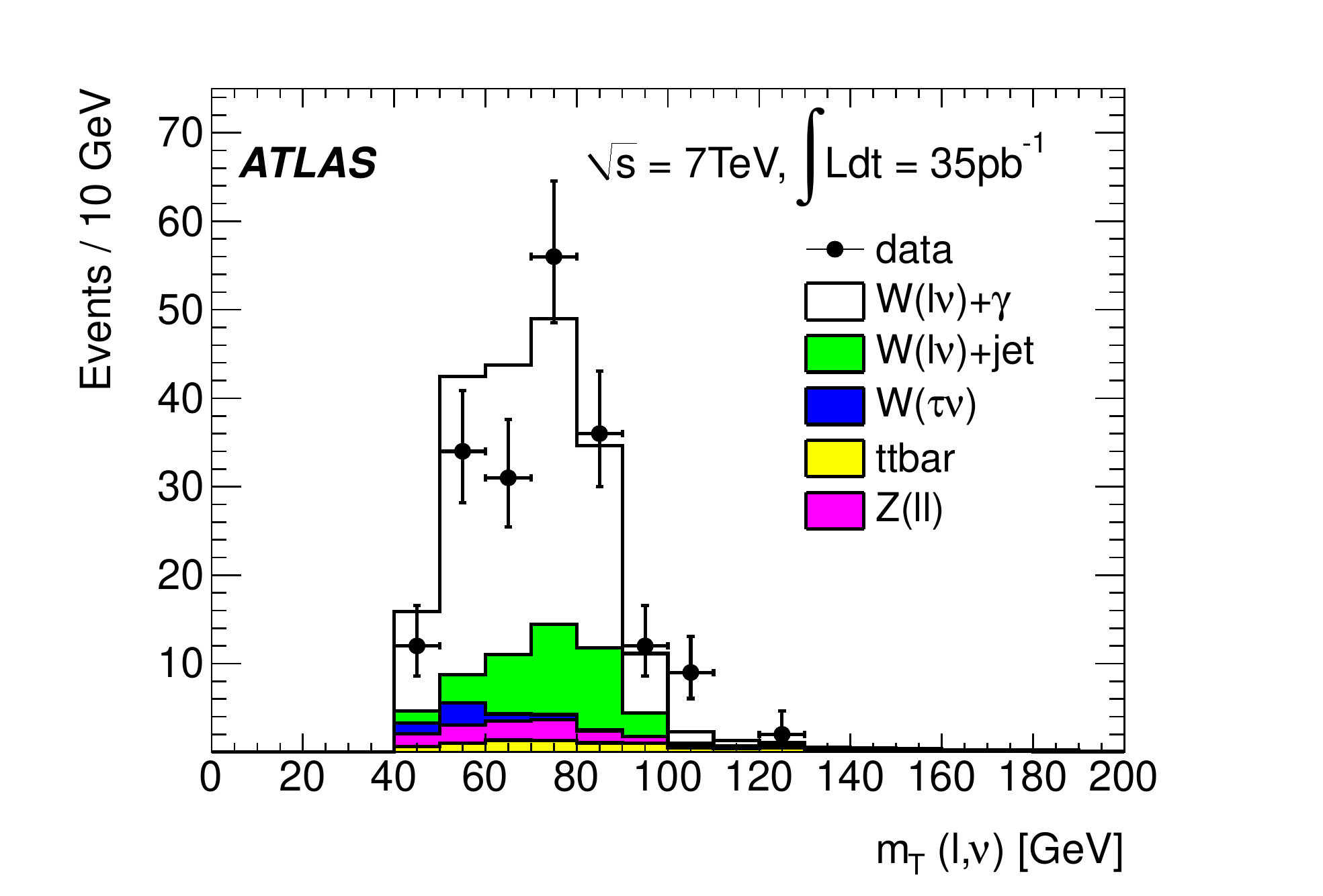}}
  \subfigure[]{\includegraphics[width=0.49\columnwidth]{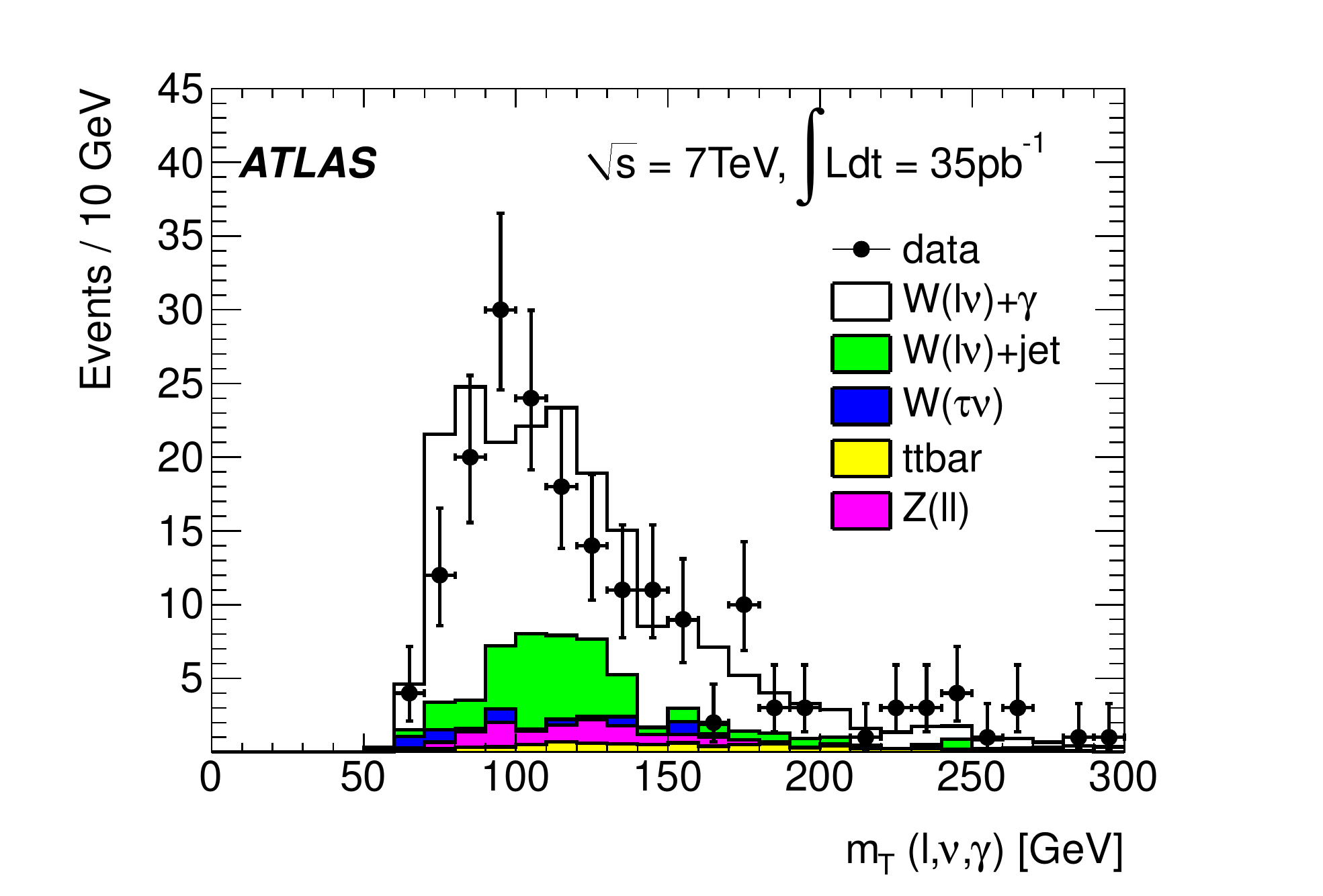}}
  \caption{Distributions for the combined electron and muon decay channels
           of the photon transverse energy (a), $\Delta R$ between lepton and photon
           (b), two body transverse mass ($m_{\mathrm{T}}(l,\nu)$) (c)
           and three body transverse mass ($m_{\mathrm{T}}(l,\nu,\gamma)$) (d) of the $W\gamma$ candidate events.
           MC predictions for signal and backgrounds are also shown.}
  \label{fig:Wg_kin}
\end{figure*}

The distributions of kinematic variables from the data are compared to signal
plus background expectations using the combined electron and muon channels
for the selected $W\gamma$ and $Z\gamma$ event candidates.
The distributions of the photon $E_{\mathrm{T}}$, $\Delta R$ between lepton
and photon, the two body transverse mass $m_{\mathrm{T}}(l,\nu)$
and the three body transverse mass $m_{\mathrm{T}}(l,\nu,\gamma)$
of $W\gamma$ candidates are shown in Fig.~\ref{fig:Wg_kin}.
The three body transverse mass, $m_{\mathrm{T}}(l,\nu,\gamma)$, is defined
in Equation (\ref{equ:MT3}) \cite{PhysRevD.48.5140}

\begin{eqnarray}
 m_{\mathrm{T}}^2(l,\nu,\gamma) &=& ( \sqrt{M^2_{l\gamma}+|\vec{p}_{\mathrm{T}}(\gamma)+\vec{p}_{\mathrm{T}}(l)|^2}+E_{\mathrm{T}}^{\mathrm{miss}})^2      \nonumber \\
   && -| \vec{p}_{\mathrm{T}}(\gamma)+\vec{p}_{\mathrm{T}}(l)+\vec{E}_{\mathrm{T}}^{\mathrm{miss}}|^2\nonumber  \\ 
\label{equ:MT3}
\end{eqnarray}

where $M_{l\gamma}$ is the invariant mass of the lepton-photon system.
In the photon distribution (Fig.~\ref{fig:Wg_kin}a) the data show a slight excess
over expectation at high $E^{\gamma}_{\mathrm{T}}$.
However the excess is not significant as there are 9 observed events for
$E^{\gamma}_{\mathrm{T}}>85$~GeV and we expect about 5 events.

\begin{figure}
  \centering
  \subfigure[]{\includegraphics[width=0.49\columnwidth]{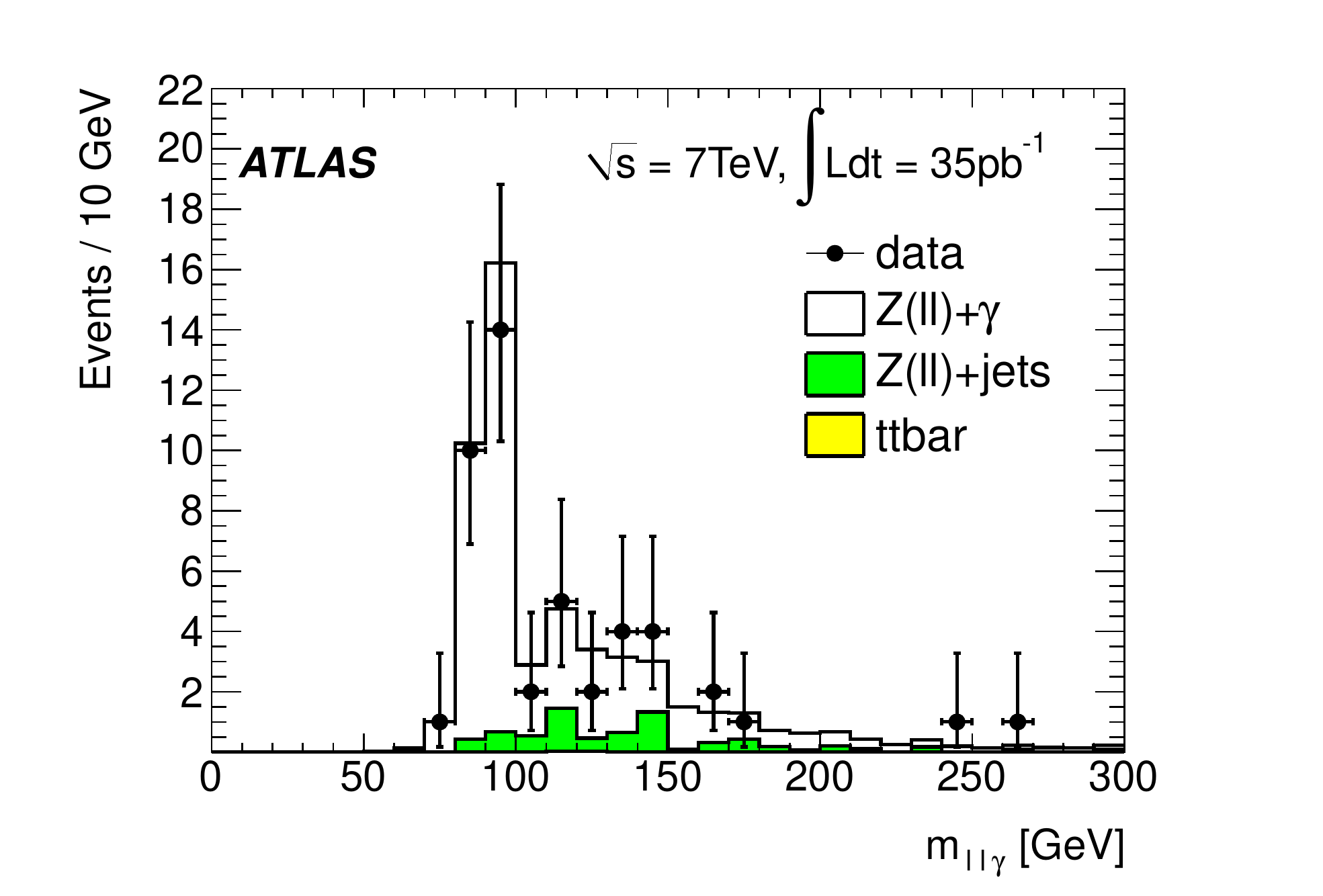}}
  \subfigure[]{\includegraphics[width=0.49\columnwidth]{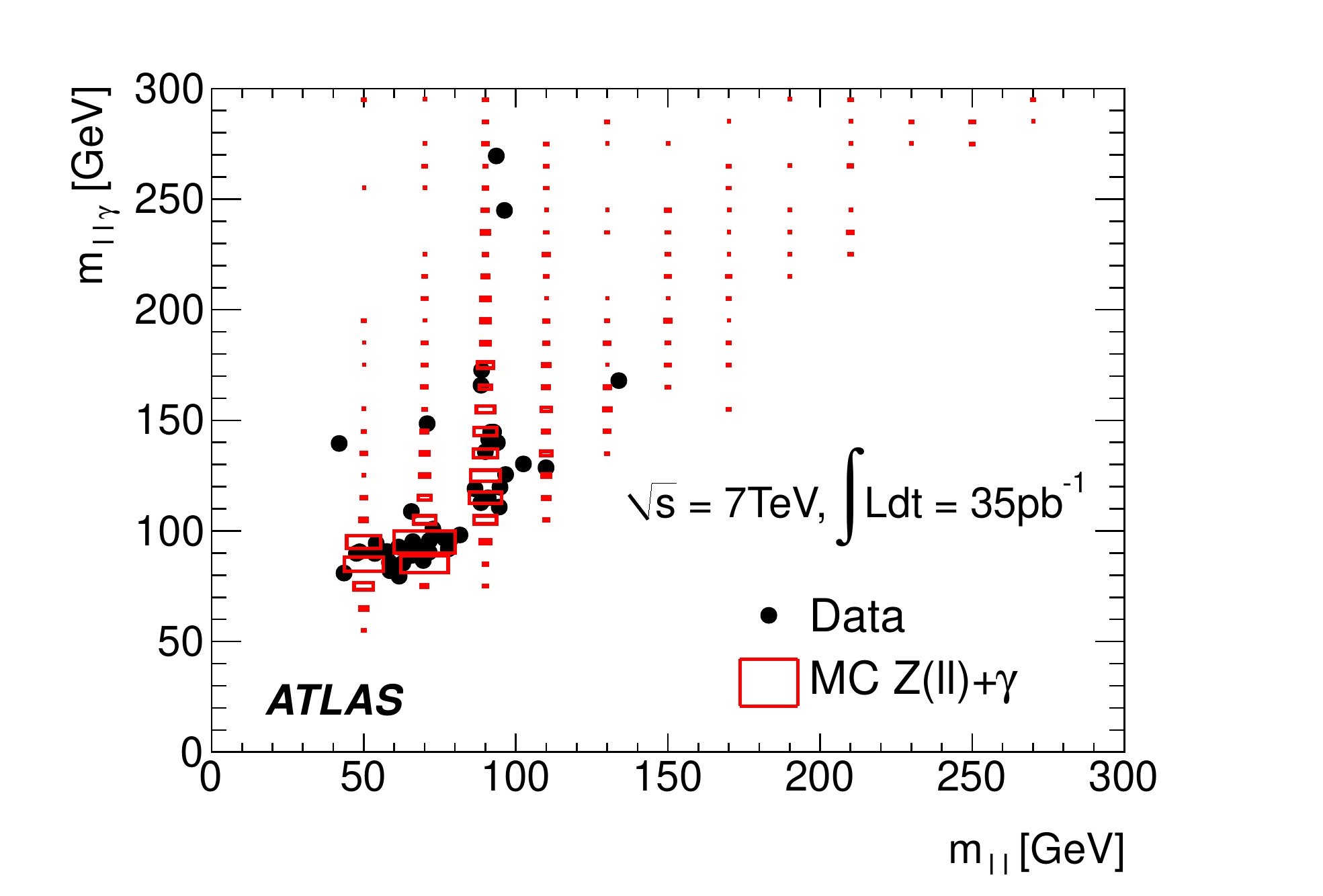}}
  \caption{(a) Three body invariant mass $m_{l^{+}l^{-}\gamma}$ distribution
           for $Z\gamma$ data candidate events. MC predictions for signal and backgrounds are also shown.
           (b) Two-dimensional plots of $m_{l^{+}l^{-}\gamma}$ vs
           $m_{l^{+}l^{-}}$ for $Z\gamma$ data candidate events.
           The MC signal prediction is also shown.
           Both the electron and muon decay channels are included.}
  \label{fig:Zg_kin}
\end{figure}

The distributions of the three body invariant mass $m_{l^{+}l^{-}\gamma}$
and the two-dimensional plots of $m_{l^{+}l^{-}\gamma}$ vs
$m_{l^{+}l^{-}}$ for the $Z\gamma$ candidates are shown in
Fig.~\ref{fig:Zg_kin}.
The data points are compared to the sum of the NLO SM predictions
for the $W\gamma$ and $Z\gamma$ plus the various background contributions.
All backgrounds, except the $W+$jets for the $W\gamma$ analysis,
are estimated from simulation and normalized with the predicted NLO cross
section values.
For the $W+$jets contribution, the shape of the background is taken
from simulations while the absolute normalization is determined from
a data-driven method described in Section~\ref{sec:background}.

%%%%%%%%%%%%%%%%%%%%%%%%%%%%%
% Figure
%%%%%%%%%%%%%%%%%%%%%%%%%%%%%

%%%%%%%%%%%%%%%%%%%%%%%%%%%%%%%%%%%%%%%%%%%%%%%%%%%
%% Efficiency
%%%%%%%%%%%%%%%%%%%%%%%%%%%%%%%%%%%%%%%%%%%%%%%%%%%
%% \input{efficiency}   % 1.5 pages

\section{Efficiency Estimation}
\label{sec:efficiency}

\subsection{Trigger Efficiency}
\label{sec:trg_eff}

The performance of the electron high $p_{\mathrm{T}}$ trigger has been measured
with data and found to be 99$\pm$1\% efficient for both ``medium'' and ``tight'' electrons
with $E_{\mathrm{T}}>$ 20 \GeV{}, with negligible $\eta$ and $E_{\mathrm{T}}$ dependence~\cite{WZpaper}.
The efficiency of the muon trigger is also measured with data, using 
$Z \rightarrow \mu^{+}\mu^{-}$ events~\cite{ATLAS_Wjet}.
The overall efficiencies to trigger on the $W\gamma$ and $Z\gamma$ events, in the
muon decay channel, are $86.2\pm0.5\%$ and $97.5\pm0.2\%$ respectively.
The electron (muon) trigger efficiency is measured with respect to an electron (muon)
candidate which has passed the offline selection cuts.
The muon trigger efficiency is lower than the electron trigger efficiency due to
limited coverage of the trigger chambers.

\subsection{Lepton Identification Efficiency}
\label{sec:lep_eff}

The electron identification efficiency $\varepsilon_{e}^{\mathrm{ID}}$ is
defined as the probability of electrons in signal events reconstructed
within the kinematic and geometric requirements to pass the
identification quality cuts~\cite{WZpaper}.  The efficiency for the
``tight'' selection in $W\gamma$ events is 73$\pm$4\%.  For the
``medium'' selection in $Z\gamma$ events, the efficiency is 92$\pm$2\%
and 87$\pm$3\% for the leading and sub-leading electron, respectively.
These efficiencies are evaluated from signal MC events with
scale factors applied to correct for discrepancies with data.  The
scale factors are obtained by comparing the electron efficiency in MC
to an \emph{in situ} electron efficiency measured in data from
unbiased probe electrons selected together with a well identified tag
electron in $Z \rightarrow e^{+}e^{-}$ candidate events, and from
unbiased probe electrons in selected $W \rightarrow e\nu$ candidate
events with large and isolated $E_{\mathrm{T}}^{\mathrm{miss}}$ recorded by the
$E_{\mathrm{T}}^{\mathrm{miss}}$ trigger.
The uncertainties on $\varepsilon_{e}^{\mathrm{ID}}$
account for background contamination in the unbiased probe electron
sample, and the potential bias from tag requirements of the \emph{in
  situ} efficiency measurement.  
The results of the two \emph{in situ}
efficiency measurements from $Z \rightarrow ee$ and $W \rightarrow
e\nu$ are combined with weights proportional to their uncertainties.

Unbiased muons from $Z\to \mu^{+}\mu^{-}$ candidate events are used to cross
check the muon identification efficiency $\varepsilon_{\mu}^{\mathrm{ID}}$
calculated with the MC signal sample~\cite{WZpaper,ATLAS_Wjet}.
The single muon identification efficiency for the
$W\gamma$ and $Z\gamma$ analyses is estimated to be $89 \pm 1 \%$.
The muon momentum scale and resolution are studied by comparing the
mass distribution of $Z\to \mu^+\mu^-$ in data and MC~\cite{WZpaper}.
The uncertainty in the acceptance of the $W\gamma$ ($Z\gamma$) signal
events due to the uncertainties in the corrections of the muon
momentum scale and resolution of the MC is $\sim 0.3\%$
($\sim 0.5\%$).

%%%%%%%%%%%%%%%%%%%%%%%%%%%%%%%%%%%%%%%%%%%%%%%%%%%
%% Photon Efficiency
%%%%%%%%%%%%%%%%%%%%%%%%%%%%%%%%%%%%%%%%%%%%%%%%%%%
%% \input{photoneff}   % 1.5 pages

\subsection{Photon Identification Efficiency}
\label{sec:ph_eff}

The photon identification efficiency, $\varepsilon_{\gamma}^{\mathrm{ID}}$, is
defined as the probability of photons in signal events, reconstructed
within the kinematic and geometric acceptance to pass the photon
identification requirements.  The photon identification efficiency is
determined from $W\gamma$ and $Z\gamma$ MC samples where the
discriminating variable distributions are corrected (by simple shifts)
to account for observed discrepancies between data and simulation.
Corrections for each discriminating variable are calculated
separately for photons in the range $|\eta|<1.8$ and
$|\eta|>1.8$. This separation is motivated by the significantly larger
discrepancies observed in the high pseudorapidity region where the
amount of material in front of the calorimeter is known less well.
The data/simulation corrections are determined by comparing
the discriminating variable distributions for photons in signal MC samples 
and candidate photons in $W\gamma$ data events
(before the isolation requirement).  The impact of the
corrections on the photon identification efficiency is -3\% (-5\%)
resulting in an estimated $\varepsilon_{\gamma}^{\mathrm{ID}}$ of 71\% (67\%)
for photons in the range $|\eta|<1.8$ ($|\eta|>1.8$).
The main source of  systematic uncertainty comes from the knowledge of the upstream material.
A dedicated simulated sample that includes additional material in the
inner detector and in front of the electromagnetic calorimeter was
used to assess the impact of a different account of material budget on
the photon identification efficiency. The resulting uncertainty on
$\varepsilon_{\gamma}^{\mathrm{ID}}$ is 6.3\% (7.5\%) for photons in the range
$|\eta|<1.8$ ($|\eta|>1.8$).
Other sources of uncertainty arise
from the simple shift approximation for the data/simulation
corrections (3\%), from the discriminating variable distribution bias
due to background contamination in the $W\gamma$ photon candidate data
sample (4\%), and from inefficiencies in the reconstruction of photon
conversions (2\%).
Since only prompt photons are present in the $W\gamma$ and $Z\gamma$
MC samples, the efficiency of the fragmentation photon
component is calculated
using an {\sc alpgen} \cite{ALPGEN} ``$W+1$ jet'' fully simulated sample by
selecting events with a high $E_{\mathrm{T}}$ photon produced in the jet
fragmentation.  The fractional contribution of fragmentation photons
to the total cross section is estimated by the Baur NLO generator (see
Section~\ref{sec:intro}) to be 8\%. 
Since there is a large uncertainty on the fragmentation photon
contribution to the $W\gamma$ and $Z\gamma$ cross sections, a
conservative error of 100\% is considered on such an estimate which  
leads to an additional 3\% uncertainty on the photon
identification efficiency.

Taking into account all the contributions, the overall uncertainty on
the photon reconstruction and identification efficiency is then
estimated to be 10.2\% (13.0\%) for photons in the range $|\eta|<1.8$
($|\eta|>1.8$).

\subsection{Photon Isolation Efficiency}
\label{sec:ph_eff_iso}

The efficiency, $\varepsilon_{\gamma}^{\mathrm{iso}}$, of the photon isolation
requirement is estimated with the signal $W\gamma$ and $Z\gamma$ MC
and cross checked with data using electrons from the $Z\to
e^+e^-$ sample (after taking into account the differences between the
electromagnetic showering of electrons and photons).
The resulting photon isolation efficiency, within its systematic
uncertainty, is found to be consistent with the one derived from the
signal MC.  The systematic uncertainties for
$\varepsilon_{\gamma}^{\mathrm{iso}}$ are due to the background contamination
in the electron sample (1\%), the shape differences of the $E^{\mathrm{iso}}_{\mathrm{T}}$
distribution between electrons and photons (0.6\%), and the
differences in $p_{\mathrm{T}}$ spectrum between electrons and photons (1.5\%).
As for the photon identification efficiency, the
 $\varepsilon_{\gamma}^{\mathrm{iso}}$ for the fragmentation components is
 obtained from an {\sc alpgen} ``$W+1$ jet'' fully simulated sample and
 an additional 3\% uncertainty is quoted to account for the
 uncertainty on the fragmentation photon contribution.  The overall
 $\varepsilon_{\gamma}^{\mathrm{iso}}$ is 95\% with a total estimated
 uncertainty of 3.3\%.

%%%%%%%%%%%%%%%%%%%%%%%%%%%%%%%%%%%%%%%%%%%%%%%%%%%
%% Background
%%%%%%%%%%%%%%%%%%%%%%%%%%%%%%%%%%%%%%%%%%%%%%%%%%%
%% \input{background}   % 2 pages

\section{Background Determination and Signal Yield}
\label{sec:background}

The dominant sources of background for this analysis are from
$W(Z)$+jets events where photons from the decay products of mesons
produced by the jet fragmentation (mainly $\pi^0 \to \gamma \gamma$)
pass the photon selection criteria. Since the fragmentation functions
of quarks and gluons into hadrons are poorly constrained by
experiments, these processes are not well modeled by $W$+jets MC simulations.
For the $W\gamma$ analysis the amount of this
background is estimated from ATLAS data while for the $Z\gamma$
analysis, due to the limited statistics, a MC based
estimation is performed and a large uncertainty of 100\% is assigned.
Additional backgrounds from other processes, such as $W \rightarrow
\tau \nu$, $t\bar t$, and $Z \rightarrow e^+e^-(\mu^+ \mu^-)$
(misidentified as $W\gamma$) for the $W\gamma$ analysis, and $t\bar t$
and $Z+$jets for the $Z\gamma$ analysis will be %HERE
%and $Z \rightarrow \tau^+ \tau^-$ for the $Z\gamma$ analysis will be %HERE
referred to collectively as ``EW+$t\bar t$ background'' and their
contribution is estimated from MC simulation.

The background from mesons decaying to photons is determined directly
from the selected $W\gamma$ events using a two-dimensional sideband
method.  This allows the extraction of the $W\gamma$ signal yield
directly from data.  Although currently limited in statistics, this
method is preferred over use of average photon background estimates
from high statistics jet trigger data samples because of the very
different probability for gluon and quark initiated jets to pass the
photon identification criteria (estimated to be different by one order
of magnitude~\cite{atlas_det}), and the poor knowledge of the quark to
gluon ratio between jets in $W$+jets events and generic inclusive jet
production.
 
\begin{figure}
  \centering
  \includegraphics[width=0.5\columnwidth]{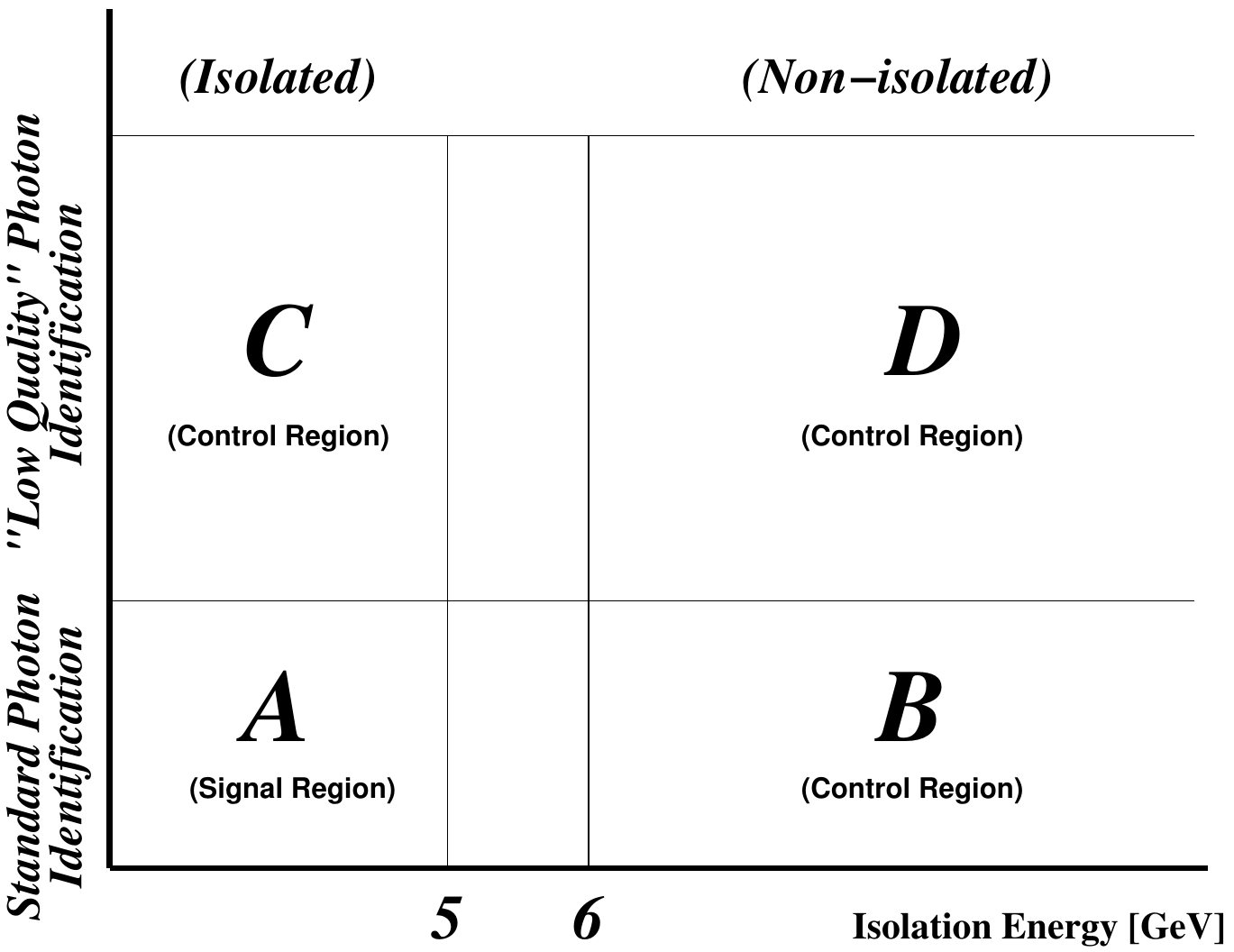}
  \caption{Sketch of the two-dimensional plane defining the 4 regions
    used in the sideband method.  Region A is the signal region. The
    non-isolated control regions (B and D) are defined for photons
    with $E^{\mathrm{iso}}_{\mathrm{T}}>6$ \GeV{}. The ``low quality photon
    identification'' control regions (C and D) include photons
    passing all the identification criteria except the strip layer
    discriminating variable requirements (see Section~\ref{sec:reco}).}
  \label{fig:tight_vs_iso}
\end{figure}

The two variables used for the sideband method are  $E^{\mathrm{iso}}_{\mathrm{T}}$ and
the identification ``quality'' of the photon candidate.
Three control regions are defined to estimate the amount of $W$+jets
background in the signal region (see Fig.~\ref{fig:tight_vs_iso}).
The signal yield of the selected $W\gamma$ sample is extracted by simply
subtracting from the number of candidate events the amount of
background in the signal region $N_{\mathrm{A}}$. This can be
determined by studying the background in the three
control regions with the assumption that for the background the ratio of
isolated to non-isolated events in the sample passing the photon
identification criteria ($N_{\mathrm{B}}$/$N_{\mathrm{A}}$) is the same as in the sample
passing the ``low quality'' identification criteria
($N_{\mathrm{D}}$/$N_{\mathrm{C}}$). Finally the  backgrounds in the control
regions are taken directly from the number of observed events in
data. Corrections are applied to subtract the contribution in these
regions from signal events (estimated from MC to be around
10\% in region C, few percent in region B, and to be negligible in region D) and the
contribution from ``EW+$t\bar t$ background'' (of the order of 10\% in
all three regions).

%%%%%%%%%%%%%%%%%%%%%%%%%%%%%
% Fig.
%%%%%%%%%%%%%%%%%%%%%%%%%%%%%

The $W+$jets background contribution as estimated by this data-driven
method is reported in Table~\ref{tab:Nobs}. In the same table the
estimated $W\gamma$ signal yield as well as the total background and
signal yield for the $Z\gamma$ analysis are shown. The effective purity, $P$,
of the $W\gamma$ ($Z\gamma$) sample, defined as the fraction of signal in the
selected events (after the subtraction of the ``EW+$t\bar t$
background'' contribution), is calculated to be around 80\%
(85\%).

\begin{table*}[!htbp]
  \centering 
  %\begin{tabular}{ccccc}
  \begin{tabular}{|c|c|c|c|c|}
    %\hline
    \hline
  Process &  Observed  & EW+$t\bar t$  &  $W+$jets   & Extracted \\ 
             & events  &  background  &   background   & signal \\ 
\hline
$N_{obs}(W\gamma\rightarrow e^{\pm}\nu\gamma)$         & 95 & $10.3 \pm 0.9 \pm 0.7$   &  $16.9 \pm 5.3 \pm 7.3$ & $67.8 \pm 9.2 \pm 7.3$\\
\hline
$N_{obs}(W\gamma \rightarrow \mu^{\pm} \nu \gamma )$   & 97 & $11.9 \pm 0.8 \pm 0.8$   &  $16.9 \pm 5.3 \pm 7.4$ & $68.2 \pm 9.3 \pm 7.4$\\

\hline
 Process  &   Observed  & \multicolumn{2}{|c|}{EW+$t\bar t$}  &Extracted \\ 
   &    events & \multicolumn{2}{|c|}{background}  & signal \\ 

\hline
$N_{obs}(Z\gamma\rightarrow e^+ e^- \gamma)$       & 25 &  \multicolumn{2}{|c|}{$3.7 \pm 3.7$} & $21.3 \pm 5.8 \pm 3.7$ \\
\hline
$N_{obs}(Z\gamma\rightarrow \mu^+ \mu^- \gamma)$   & 23 &  \multicolumn{2}{|c|}{$3.3 \pm 3.3$} & $19.7 \pm 4.8 \pm 3.3$ \\
\hline
%\hline
  \end{tabular} 
  \caption{Numbers of the total observed candidate events, estimated
number of background and estimated number of signal events for the
$pp\rightarrow l^{\pm} \nu\gamma + X$ and $pp \rightarrow l^+l^-\gamma + X$
selected samples.
Where two uncertainties are quoted the first is statistical and the
second represents an estimate of systematics.
Statistical errors in MC predictions are treated as a systematic
in the propagation of uncertainties on the W+jets background and the
extracted signal.
The $W+$jets background contribution is estimated from ATLAS data with a
two-dimensional sideband method.
For the $pp \rightarrow l^+l^-\gamma + X$ process the uncertainty on
the MC based background estimate is 100\%.}
  \label{tab:Nobs} 
\end{table*}

The accuracy of the $W$+jets background determination with the
two-dimensional sideband method has been carefully assessed.  The
uncertainty related to the definition of the control regions is
determined by studying the impact of possible variations of their
definitions. For the non-isolated control regions (B and D) the lower
boundary of 6 GeV has been shifted by $\pm 1$ GeV, probing different
mixtures of background and $W\gamma$ signal event contamination.  For
the ``low quality'' photon identification control regions (C and D)
two alternative choices of strip layer discriminating variable
criteria are tested. These changes of control region
definitions lead to respectively a 4\% and a 9\% variation of the
effective purity estimate.  The contamination from $W\gamma$ signal events in
the control regions is strongly correlated with the photon
identification efficiency in the signal region (an overestimate of the
latter induces an underestimate of the former). Shifting the
discriminating variable distributions of the signal MC in a
way similar to the one described in Section~\ref{sec:ph_eff} results
in an impact on the effective purity estimation of the order of 3\%.
Finally, the accuracy on the assumption that the correlations between the
two-dimension variables (namely the energy isolation and the photon
identification quantities) are negligible for background events has
been evaluated by applying the same method to background samples
extracted from $W$+jets MC events. The corresponding purities
are all found to be compatible with zero and their values are used to
determine the systematic uncertainty associated to the method,
estimated to be 3\%.  For the ``EW+$t\bar t$ background'' estimation,
the corresponding NLO theoretical cross section uncertainty (between
6\% to 7\% depending on the process) and the luminosity uncertainty
(3.4\%) are used.

\begin{figure}
  \centering
  \subfigure[]{\includegraphics[width=0.49\columnwidth]{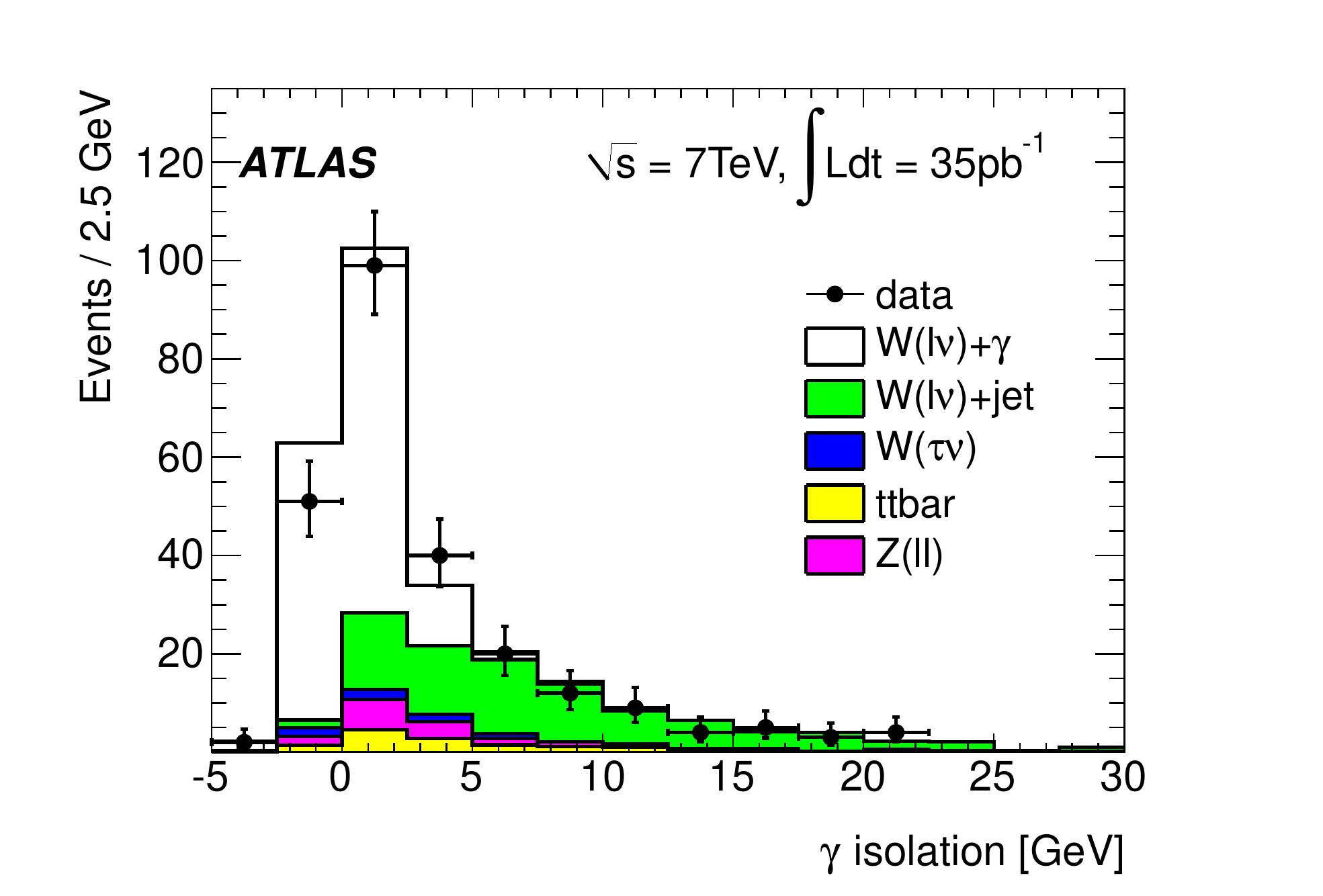}}
  \subfigure[]{\includegraphics[width=0.49\columnwidth]{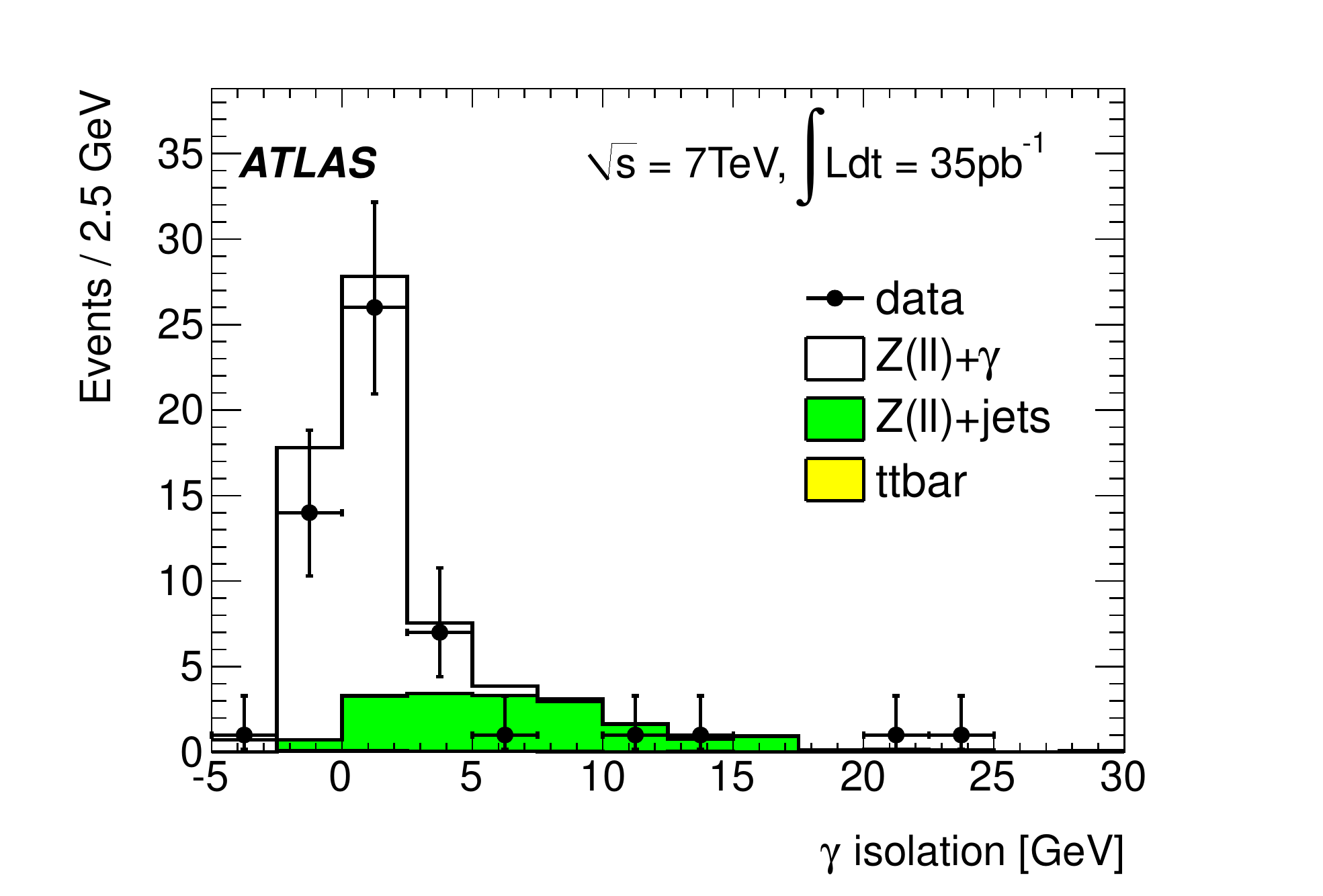}}
  \caption{Photon isolation distribution for photon candidates in the
    $W\gamma$ (a) and in the $Z\gamma$ (b) data events (points).
    The shape of the predicted $W+$jets background is taken
    from the data photon isolation distribution of events in the
    control regions C-D while the normalization is
    determined by the two-dimensional sideband data-driven method. The
    predicted contributions from the other backgrounds and from
    the signal are taken from MC.}
  \label{fig:Wg_iso}
\end{figure}

In Fig.~\ref{fig:Wg_iso}a (\ref{fig:Wg_iso}b), the
$E^{\mathrm{iso}}_{\mathrm{T}}$ distribution of photon candidate events
in the $W\gamma$ ($Z\gamma$) combined 
sample is shown along with the predicted contributions for the
background. 

%%%%%%%%%%%%%%%%%%%%%%%%%%%%%
% Figure
%%%%%%%%%%%%%%%%%%%%%%%%%%%%%

%%%%%%%%%%%%%%%%%%%%%%%%%%%%%%%%%%%%%%%%%%%%%%%%%%%
%% Cross Section + Results
%%%%%%%%%%%%%%%%%%%%%%%%%%%%%%%%%%%%%%%%%%%%%%%%%%%
%% \input{cs}           % 4 pages

\section{Cross Section Measurements and Comparison to Theoretical Calculations}
\label{sec:cs}

\subsection{Fiducial cross section measurement for $W \gamma$ and $Z \gamma$}
\label{sec:fid_cs}

The measurements for the fiducial cross sections for the processes
$pp \to l^{\pm}\nu\gamma+X$ and $pp \to l^+l^-\gamma+X$ can be expressed as

\begin{equation}
\sigma_{pp \rightarrow l^{\pm}\nu\gamma(l^{+}l^{-}\gamma)}^{\mathrm{fid}} = \frac{N^{\mathrm{sig}}_{W\gamma(Z\gamma)}} {C_{W \gamma(Z\gamma)} \cdot L_{W\gamma(Z\gamma)}}
\label{Equ:cs_fid}
\end{equation}
where 
\begin{itemize}
\item $N_{W\gamma}^{\mathrm{sig}}$ and $N_{Z\gamma}^{\mathrm{sig}}$ denote the number
  of background-subtracted signal events passing the selection
  criteria of the analyses in the $W \gamma$ and $Z \gamma$ channels.
  The $N^{\mathrm{sig}}$ values for both $W\gamma$ and $Z\gamma$ processes are given in
  Table~\ref{tab:Nobs}.
\item $L_{W\gamma}$ and $L_{Z\gamma}$ denote the integrated
  luminosities for the channels of interest.
\item $C_{W\gamma}$ and $C_{Z\gamma}$ are correction factors and denote the probability for
events generated within the fiducial region of the phase-space (as defined in
Table~\ref{tab:fiducialcut}) to pass the final selection
requirements.
\end{itemize}

\begingroup
\begin{table*}[!htbp]
  \centering 
  \begin{tabular}{|l|c|c|c|c|}
    \hline
 \multicolumn{5}{|c|}{{\bf Fiducial phase space}}  \\
\hline
   &  $e^{\pm}\nu\gamma$ &  $e^+ e^- \gamma$ & $\mu^{\pm}\nu\gamma$     & $\mu^+ \mu^- \gamma$\\ 
\hline
 $E_\mathrm{T}^l$($p_\mathrm{T}^l$) & $E_\mathrm{T}^e>20 \GeV{}$ & $E_\mathrm{T}^e>20\GeV{}$ & $p_\mathrm{T}^\mu>20\GeV{}$  & $p_\mathrm{T}^\mu>20\GeV{}$ \\
                        & $p_\mathrm{T}^\nu>25\GeV{}$   & - & $p_\mathrm{T}^\nu>25\GeV{}$ & - \\
\hline
$\eta_l$  & $0<|\eta_e|<1.37$& $0<|\eta_e|<1.37$ & $|\eta_\mu|<2.4$& $|\eta_\mu|<2.4$  \\
 & or  & or        &         & \\
 & $1.52<|\eta_e|<2.47$& $1.52<|\eta_e|<2.47$ &   & \\
\hline
Boson cut & $m_\mathrm{T}>40$ \GeV{}  & $m_{ee}>40$ \GeV{} & $m_\mathrm{T}>40$ \GeV{}  & $m_{\mu\mu}>40$ \GeV{} \\
\hline
 & \multicolumn{4}{|c|}{$E_\mathrm{T}^\gamma>15$  \GeV{}}\\
Photon & \multicolumn{4}{|c|}{$0<|\eta_\gamma|<1.37$ or $1.52<|\eta_\gamma|<2.37$}\\
 & \multicolumn{4}{|c|}{$\Delta R(l,\gamma)>0.7$ }  \\
 & \multicolumn{4}{|c|}{ $\epsilon_h^p < 0.5$ }  \\
\hline
 \multicolumn{5}{|c|}{{\bf  Phase space for production cross section}}  \\
\hline
   &  $e^{\pm}\nu\gamma$ &  $e^+ e^- \gamma$ & $\mu^{\pm}\nu\gamma$     & $\mu^+ \mu^- \gamma$\\ 
\hline
Boson   &             & $m_{ee}>40$ \GeV{} &                 & $m_{\mu\mu}>40$ \GeV{} \\
\hline
 & \multicolumn{4}{|c|}{$E_\mathrm{T}^\gamma>15$  \GeV{}}\\
Photon  & \multicolumn{4}{|c|}{$\Delta R(l,\gamma)>0.7$ }  \\
 & \multicolumn{4}{|c|}{$\epsilon_h^p < 0.5$ }  \\

\hline
  \end{tabular} 
  \caption{Definition of the fiducial
  phase space at the particle level, where the measurements are performed and the extended phase space
  (common to all measurements), where the production cross sections are
  evaluated. $\epsilon_h^p$ is defined in Section ~\ref{sec:theory_xsection}.}
\label{tab:fiducialcut}
\end{table*}

The correction factors $C_{W\gamma(Z\gamma)}$  include all trigger efficiencies, selection efficiencies and reconstruction efficiencies of the photon and leptons. 

\begin{equation}
C_{W\gamma}=\varepsilon_{\mathrm{event}}^{W\gamma} \cdot \varepsilon_{\mathrm{lep}}^{\mathrm{ID}} \cdot \varepsilon_{\mathrm{trig}}^{W\gamma} \cdot \varepsilon^{{\rm ID}}_\gamma \cdot \varepsilon_\gamma^{{\rm iso}}  \cdot \alpha_{\mathrm{reco}}^{W\gamma}
\end{equation}
\begin{equation}
C_{Z\gamma}=\varepsilon_{\mathrm{event}}^{Z\gamma} \cdot (\varepsilon_{\mathrm{lep}}^{\mathrm{ID}})^2 \cdot \varepsilon_{\mathrm{trig}}^{Z\gamma} \cdot \varepsilon^{{\rm ID}}_\gamma \cdot \varepsilon_\gamma^{{\rm iso}}  \cdot \alpha_{\mathrm{reco}}^{Z\gamma}
\end{equation}
where
\begin{itemize}
\item $\varepsilon_{\mathrm{trig}}^{W\gamma}$ and $\varepsilon_{\mathrm{trig}}^{Z\gamma}$ denote the probability of $W\gamma$ and $Z\gamma$ events to be recorded by the electron or muon trigger.

\item $\varepsilon_{\mathrm{event}}^{W\gamma}$ and $\varepsilon_{\mathrm{event}}^{Z\gamma}$ denote event selection efficiencies (including efficiency of primary vertex requirement).

\item $\varepsilon_{\mathrm{lep}}^{\mathrm{ID}} $ denotes lepton identification efficiency.
\item $\varepsilon^{{\rm ID}}_\gamma$ denotes photon identification efficiency.
\item $\varepsilon^{{\rm iso}}_\gamma$ denotes photon isolation efficiency.
\item $\alpha_{\mathrm{reco}}^{W\gamma}$ and $\alpha_{\mathrm{reco}}^{Z\gamma}$
account for all differences observed between the efficiencies of applying the
kinematic and geometrical cuts at generator level and reconstruction level.
Their values are not closed to 100\% mainly due to
acceptance loss of the electron and photon reconstruction caused by some
inoperative readouts in the electromagnetic calorimeter, reconstruction
efficiencies of the leptons and photon, and the detector resolution on the
lepton transverse momenta/energies and on the missing transverse energy.
\end{itemize}

The central values of the correction factors $C_{W\gamma}$ and
$C_{Z\gamma}$ are computed using $W\gamma$ and $Z\gamma$ signal
MC samples, with scale factor corrections to account for discrepancies in trigger, lepton and photon
selection efficiencies between data and MC, as described in
Section~\ref{sec:efficiency}. The central values of the correction factors $C_{W\gamma}$ ($C_{Z\gamma}$) of both electron and muon channels together with their components are given in Table~\ref{tab:input_wg}.

The breakdown of the uncertainties on $C_{W\gamma}$ and $C_{Z\gamma}$ is reported
in Table~\ref{tab:sys_CwgElec} and \ref{tab:sys_CwgMuon}.
The uncertainties related to the efficiency components of
$C_{W\gamma}$ and $C_{Z\gamma}$ have been discussed in
Section~\ref{sec:efficiency}.
 Other sources of uncertainties include: 
\begin{itemize}

\item
The impact of the EM energy scale uncertainty is evaluated by propagating the EM energy scale uncertainties to the number of accepted $W\gamma$ and $Z\gamma$ events. The EM energy scale uncertainty, after applying \emph{in situ} data driven calibration to correct for cluster energies of photon and electron clusters, is quoted to be 1\% in the barrel region, and 3\% in the endcap region.

\item The muon momentum scale and resolution are studied by comparing the
mass distribution of $Z\rightarrow\mu^+\mu^-$ in data and MC simulations~\cite{WZpaper}.
The uncertainty in the acceptance of the $W\gamma$ ($Z\gamma$) signal
events due to the uncertainties in the corrections of the muon
momentum scale and resolution of the MC simulations is $\sim 0.3\%$
($\sim 0.5\%$).

\item The acceptance loss from a few inoperative optical links of the calorimeter readout is evaluated from the signal MC. The imperfect modeling of this acceptance loss need to be considered in the systematics uncertainty of $C_{W\gamma}$ and $C_{Z\gamma}$. This uncertainty is estimated to be about 0.7\% for a single ($e$/$\gamma$) object.

\item  The experimental uncertainty arising
from the transport of low-energy bremsstrahlung photons through the detector material
and the response of the electromagnetic calorimeter is
estimated to be less than $0.3\%$~\cite{WZpaper}.

\item The main uncertainty on the scale of the missing transverse energy
is determined from a variation of the response of cells in topological clusters.
Other sources of uncertainty, namely the imperfect modelling of the
overall $E_\mathrm{T}^{\mathrm{miss}}$ response (e.g. from low energy hadrons)
and resolution, of the underlying event and pile-up effects are also considered.
The overall impact on $C_{W\gamma}$ is 2\%~\cite{WZpaper}.
\end{itemize}

All the quantities needed to calculate the cross sections defined in
Equation (\ref{Equ:cs_fid}), along with their uncertainties, are tabulated
in Table~\ref{tab:input_totalcs}. Using these numbers, the measured fiducial
cross sections for the $pp \rightarrow l^{\pm}\nu\gamma+X$ and
$pp \rightarrow l^+l^-\gamma+X$ processes are determined.
The results are presented in Table~\ref{tab:fidutotal_cs} and also
illustrated in Fig.~\ref{fig:totalcs}.
MC statistical uncertainties are included as part of the
cross sections systematics.
The most significant systematic uncertainties in both measurements arise
from the background estimation and the efficiencies of photon identification
and isolation.

%%%%%%%%%%%%%%%%%%%%%%%%%%%%%%%%%%%%%%%
% TABLE
%%%%%%%%%%%%%%%%%%%%%%%%%%%%%%%%%%%%%%%
\begin{table}
  \centering 
  \begin{tabular}{|c|c|c|c|c|}
    \hline
   & $pp \rightarrow e^{\pm} \nu\gamma $ &  $pp \rightarrow \mu^{\pm} \nu\gamma $ & $pp \rightarrow e^+e^-\gamma $ &   $ pp \rightarrow \mu^+\mu^-\gamma $ \\
    \hline
$\varepsilon_{\mathrm{event}}$   &  100\% & 100\% &  100\% & 100\%  \\

$\varepsilon_{\mathrm{trig}}^{\mathrm{event}}$    & 99\%      & 86\%  & 100\%   & 98\%  \\

$\varepsilon_{\mathrm{lep}}^{\mathrm{ID}}$     & 73\%      & 89\% & 90\%   & 88\%  \\

$\varepsilon_{\gamma}^{\mathrm{ID}}$   & 70\%         & 71\%  & 70\%    & 70\%    \\

$\varepsilon_{\gamma}^{\mathrm{iso}}$  & 95\%       & 96\% & 96\%    & 96\%    \\

$\alpha_{\mathrm{reco}}$         & 75\%      & 87\%     & 53\%     & 85\%    \\
\hline
$C_{V\gamma}$           & 36\%      & 46\%    & 28\%       & 43\%  \\ 

\hline
  \end{tabular} 
  \caption{Efficiency factors per lepton and $\alpha_{\mathrm{reco}}$, which enter the calculation of the
    correction factors $C_{V\gamma}$ (where $V$ denotes \emph{W} or \emph{Z} boson) for both lepton channels.  The
    trigger efficiencies are measured from data. The other
    efficiencies are determined from
    MC simulation and have been validated with data, as
    described in Section~\ref{sec:efficiency}. A detailed summary of the various
    contributions entering the uncertainty on $C_{V\gamma}$ is given
    in Table~\ref{tab:sys_CwgElec} and \ref{tab:sys_CwgMuon}.}
  \label{tab:input_wg} 
\end{table}

%%%%%%%%%%%%%%%%%%%%%%%%%%%%%%%%%%%%%%%
% TABLE
%%%%%%%%%%%%%%%%%%%%%%%%%%%%%%%%%%%%%%%
\begin{table*}
  \centering 
  \begin{tabular}{|c|c|c|c|}
    \hline
  Parameter &  $\frac{\delta C_{W\gamma}}{C_{W\gamma}}$ & $\frac{\delta C_{Z\gamma}}{C_{Z\gamma}}$ & $\delta(\frac{ C_{W\gamma}}{C_{Z\gamma}})/\frac{ C_{W\gamma}}{C_{Z\gamma}}$ \\
\hline
Channel   &   $e^{\pm}\nu\gamma$ & $e^+e^-\gamma$& Electron \\
\hline
    Trigger efficiency  & 1\% & 0.02\% & 1\% \\
    Electron  efficiency  & 4.5\%  & 4.5\% & 4.5\%  \\
    Photon  efficiency  & 10.1\%  & 10.1\% & -  \\
    EM scale and resolution  & 3\%  & 4.5\% & 1.5\% \\
    $E_\mathrm{T}^{\mathrm{miss}}$ scale and resolution  & 2\%  & -  & 2\% \\
    Inoperative readout modeling   & 1.4\%   & 2.1\% & 0.7\%  \\
    Photon simulation modeling   & 0.3\%   & 0.3\%  & 0.3\% \\
    Photon isolation efficiency  & 3.3\%   & 3.3\% & -  \\
\hline
    Total uncertainty  & 12.1\%   & 12.5\% & 5.3\%  \\
\hline
  \end{tabular} 
  \caption{Summary of the different terms contributing to the
  uncertainty on $C_{W\gamma}$ and $C_{Z\gamma}$ for the electron final
  state. The decomposition has been made such that correlations
  between the various contributions are negligible.}
  \label{tab:sys_CwgElec}
\end{table*}

%%%%%%%%%%%%%%%%%%%%%%%%%%%%%%%%%%%%%%%
% TABLE
%%%%%%%%%%%%%%%%%%%%%%%%%%%%%%%%%%%%%%%
\begin{table*}
  \centering 
  \begin{tabular}{|c|c|c|c|}
    \hline
  Parameter &  $\frac{\delta C_{W\gamma}}{C_{W\gamma}}$ & $\frac{\delta C_{Z\gamma}}{C_{Z\gamma}}$ & $\delta(\frac{ C_{W\gamma}}{C_{Z\gamma}})/\frac{ C_{W\gamma}}{C_{Z\gamma}}$  \\
\hline
Channel   &   $\mu^{\pm}\nu\gamma$ & $\mu^+\mu^-\gamma$& Muon  \\
\hline
    Trigger efficiency    & 0.6\%  & 0.2\% & 0.6\% \\
    Muon  efficiency    & 0.5\%  & 1\% & 0.5\%  \\ 
    Muon isolation efficiency   & 1\%    & 2\%  & 1\% \\ 
    Momentum scale and resolution                        & 0.3\%  & 0.5\% &  0.2\%\\
    Photon  efficiency  & 10.1\%    & 10.1\% & -  \\
    EM scale and resolution                       & 4\%    & 3\%  & 1\% \\
    $E_\mathrm{T}^{\mathrm{miss}}$ scale and resolution   & 2\%    & -  & 2\% \\
    Inoperative readout modeling    & 0.7\%  & 0.7\% & -  \\
    Photon simulation modeling   & 0.3\%  & 0.3\%  & 0.3\%\\
    Photon isolation efficiency  & 3.3\%    & 3.3\%  & - \\
\hline
    Total uncertainty            & 11.6\% & 11.2\% & 2.6\%  \\
\hline
  \end{tabular} 
  \caption{Summary of the different terms contributing to the
  uncertainty on $C_{W\gamma}$ and $C_{Z\gamma}$ for the muon final
  state. The decomposition has been made such that correlations
  between the various contributions are negligible.}
  \label{tab:sys_CwgMuon}
\end{table*}

%%%%%%%%%%%%%%%%%%%%%%%%%%%%%%%%%%%%%%%
% TABLE
%%%%%%%%%%%%%%%%%%%%%%%%%%%%%%%%%%%%%%%
\begin{table}
  \centering 
  \begin{tabular}{|c|c|c|c|c|}
\hline
   &  Central & Statistical & Systematic & Luminosity \\
   &  value &  uncertainty & uncertainty & uncertainty \\
\hline
   & \multicolumn{4}{|c|}{ $pp \rightarrow  e^{\pm} \nu\gamma $} \\
\hline
$N^{\mathrm{sig}}_{W\gamma}$   &  67.8 & 9.2 & 7.3 &-   \\
$L_{W\gamma}[\rm{pb}^{-1}]$  & 35.1 & -  & - &  1.2   \\
$C_{W\gamma}$  & 0.359  & 0.010   & 0.043 &  -  \\
$A_{W\gamma}$   & 0.131  &  0.001  &  0.006 & -  \\
\hline
   & \multicolumn{4}{|c|}{ $pp \rightarrow  e^+ e^- \gamma $} \\
\hline
$N^{\mathrm{sig}}_{Z\gamma}$   &  21.3 & 5.8 & 3.7 & -   \\

$L_{Z\gamma}[\rm{pb}^{-1}]$  & 35.1 & -  & - &  1.2   \\

$C_{Z\gamma}$  & 0.280     & 0.010  & 0.035 &  -  \\
$A_{Z\gamma}$   & 0.220  &  0.002  &  0.015 & -  \\
\hline
   & \multicolumn{4}{|c|}{ $pp \rightarrow  \mu^{\pm} \nu\gamma $} \\
\hline

$N^{\mathrm{sig}}_{W\gamma}$   &  68.2 & 9.3 & 7.4 & -  \\
$L_{W\gamma}[\rm{pb}^{-1}]$  & 33.9 & -  & - &  1.2   \\
$C_{W\gamma}$  & 0.455 & 0.010  & 0.053 &  -  \\
$A_{W\gamma}$  & 0.134   &  0.001 &  0.006 & -  \\
\hline
   & \multicolumn{4}{|c|}{ $pp \rightarrow  \mu^+ \mu^-\gamma $} \\
\hline
$N^{\mathrm{sig}}_{Z\gamma}$   &  19.7 & 4.8 & 3.3 & -  \\
$L_{Z\gamma}[\rm{pb}^{-1}]$  & 33.9 & -  & - &  1.2   \\
$C_{Z\gamma}$  & 0.429 & 0.010  & 0.048 &  -  \\
$A_{Z\gamma}$   & 0.242   &  0.002 & 0.016 & -  \\
\hline
  \end{tabular} 
  \caption{Summary of input quantities for the calculation of the
$W\gamma$ and $Z\gamma$ fiducial and production cross sections.
For each channel, the observed numbers of signal events after
background subtraction, the correction factors $C_{W\gamma(Z\gamma)}$,
the acceptance factors $A_{W\gamma(Z\gamma)}$
(see Section ~\ref{sec:mes_tot_cs}), and the
integrated luminosities are given, with their statistical,
systematic, and luminosity uncertainties.
For $C_{W\gamma(Z\gamma)}$ and $A_{W\gamma(Z\gamma)}$, the statistical
uncertainty reflects the limited statistic of the signal MC samples.}
  \label{tab:input_totalcs} 
\end{table}

\subsection{Production Cross Section Measurement for $W \gamma$ and $Z \gamma$}
\label{sec:mes_tot_cs}

The production cross sections for the $W \gamma$ and $Z \gamma$ processes
are defined for the full decay phase space of the $W$ and $Z$ bosons
and for photons with $E_{\mathrm{T}}^{\gamma}>15$ \GeV{}, $\Delta R(l,\gamma)>0.7$ and $\epsilon_h^p<0.5$.
These cross sections can be derived from fiducial cross sections by extrapolation from the fiducial
phase space to the extended phase space, where production cross sections are defined.
The definition of the production cross sections is shown in Equation~(\ref{eqn:total}).
\begin{equation}
\sigma_{pp\rightarrow l^{\pm}\nu\gamma(pp\rightarrow l^{+}l^{-}\gamma)}=
\frac{\sigma_{pp\rightarrow l^{\pm}\nu\gamma(pp\rightarrow
    l^{+}l^{-}\gamma)}^{\mathrm{fid}}}{ A_{W\gamma(Z\gamma)}}
\label{eqn:total}
\end{equation}

The acceptance factors $A_{W\gamma}$ and $A_{Z\gamma}$ are defined as the fraction of weighted events in the $W(Z)+\gamma$ LO MC sample, generated within the phase space of the production cross section, that satisfy the geometrical and kinematic constraints of the fiducial cross section as shown in Table~\ref{tab:fiducialcut}. The weight of the LO MC events is from QCD NLO correction $k$-factors, which also include contributions from fragmentation components as described in section~\ref{sec:theory_xsection}.

The systematic uncertainties on the acceptances are dominated by the
limited knowledge of the proton PDFs. These are evaluated by comparing
the acceptances obtained by adopting different PDF sets (including
CTEQ6L1 \cite{CTEQ6l1}, HERAPDF1.0 \cite{H1} and MRST LO*
\cite{pdfmrst}). Other contributions are the uncertainties due to the
NLO correction of $W\gamma$ and $Z\gamma$ production, which is derived
from the difference between the Born level acceptance and acceptance
in Baur NLO simulations.  The overall relative systematic uncertainty
on $A_{W \gamma}$ ($A_{Z \gamma}$) is 4.5\% (6.7\%), the relative
systematic uncertainty for the $A_{W \gamma}/A_{Z \gamma}$ ratio is 4\%.

The measured production cross sections for the $pp \to e^{\pm}\nu\gamma+X$, $pp \to \mu^{\pm}\nu\gamma+X$, $pp \to
e^+e^-\gamma+X$ and $pp \to \mu^+\mu^-\gamma+X$ processes are summarized in Table~\ref{tab:fidutotal_cs}.

Assuming lepton universality for the $W$ and $Z$-boson decays,
the measured cross sections in the two channels can be combined to reduce the statistical uncertainty.
The combination of electron and muon channels in the production cross section measurement is based on the assumption that the uncertainties on the integrated luminosity, on the acceptance correction factors, on the background estimation, and on photon reconstruction, identification, and isolation efficiency are fully correlated. All systematic uncertainties related to lepton efficiencies (i.e. trigger and lepton identification efficiencies) are uncorrelated.
The resulting total cross sections for $pp \to l^{\pm}\nu\gamma+X$ and $pp \to l^+l^-\gamma+X$ processes using the combined electron and muon channels are summarized in Table~\ref{tab:fidutotal_cs} and plotted in Fig.~\ref{fig:totalcs} with a comparison to SM predictions.

\subsection{The Ratio of the $W\gamma$ to $Z\gamma$ Cross Sections}
 The ratio of the $W\gamma$ to $Z\gamma$ cross sections, as defined in Equation~(\ref{Equ:R}), can be measured with a higher relative precision than the individual cross sections since both experimental and theoretical uncertainties partially cancel.
This ratio is a test of the $WW\gamma$ triple gauge coupling predicted by the SM.
\begin{equation}
R=\frac{\sigma_{pp \to l^{\pm}\nu\gamma}}{\sigma_{pp \to l^+l^-\gamma}}
\label{Equ:R}
\end{equation}
In terms of the experimental quantities defined in the previous sections, the ratio R can be written as:
\begin{equation}
 R = \frac{N^{\mathrm{sig}}_{W\gamma}}{N^{\mathrm{sig}}_{Z\gamma}} \cdot \frac{C_{Z\gamma}}{C_{W\gamma}}\cdot \frac{A_{Z\gamma}}{A_{W\gamma}}
\end{equation}

The uncertainty on the ratio of the correction factors $\frac{C_{Z\gamma}}{C_{W\gamma}}$ is evaluated separately for the electron and the muon channels, as shown in Table~\ref{tab:sys_CwgElec} and~\ref{tab:sys_CwgMuon}. The uncertainties on the ratio
of the acceptance factors $\frac{A_{Z\gamma}}{A_{W\gamma}}$ have already been discussed in Section \ref{sec:mes_tot_cs}. The uncertainties on $N^{\mathrm{sig}}_{W\gamma}$ and $N^{\mathrm{sig}}_{Z\gamma}$, as shown in Table~\ref{tab:Nobs}, are considered as uncorrelated in the ratio measurement.
The measured ratios \emph{R} in the fiducial phase space and in the total phase space are shown in Table~\ref{tab:tableratio} and also illustrated in Fig.~\ref{fig:figratio}.

\subsection{Comparison to Theoretical Calculation}
\label{sec:theory}

\begin{figure}
  \centering
\includegraphics[width=0.60\columnwidth]{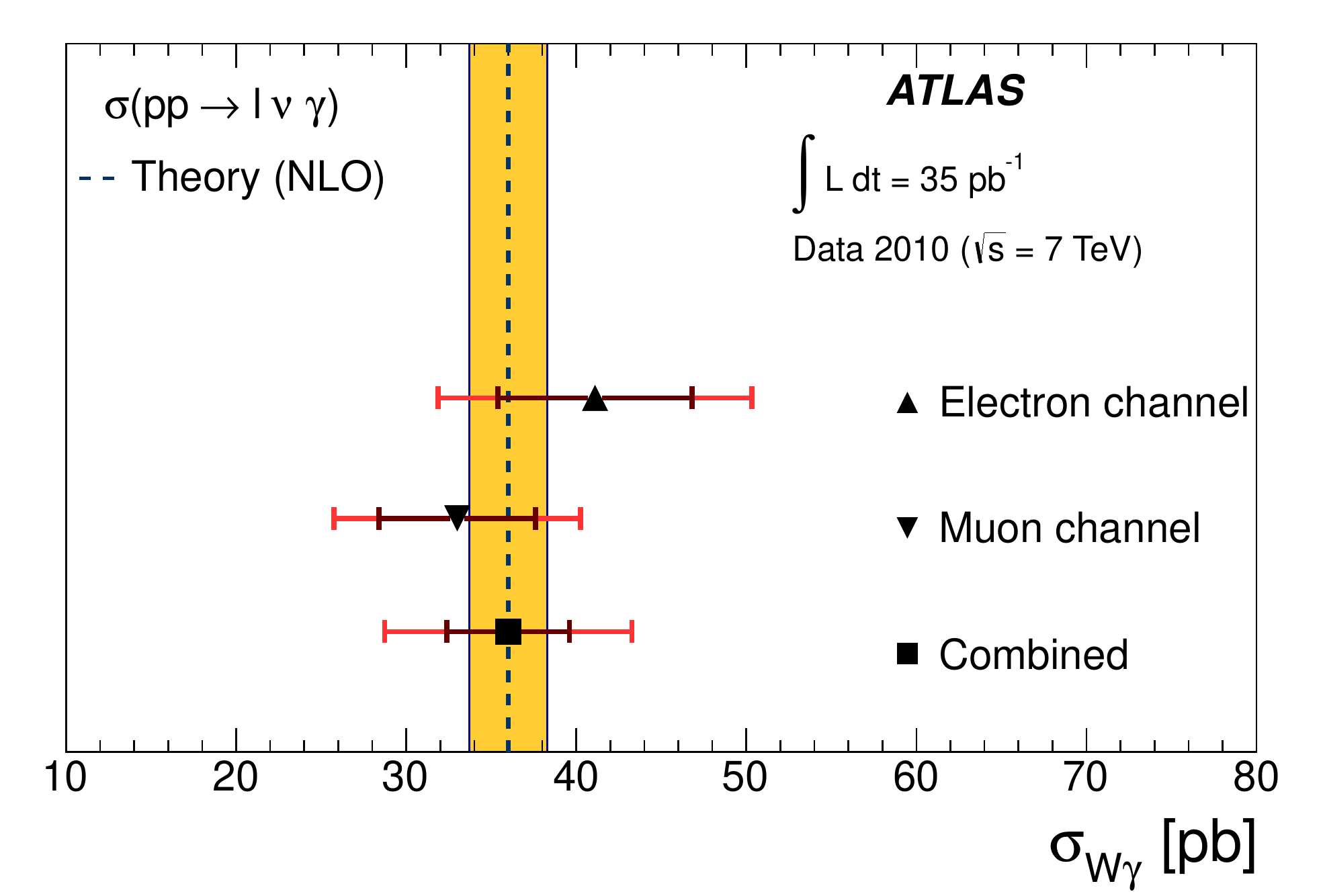}
\includegraphics[width=0.60\columnwidth]{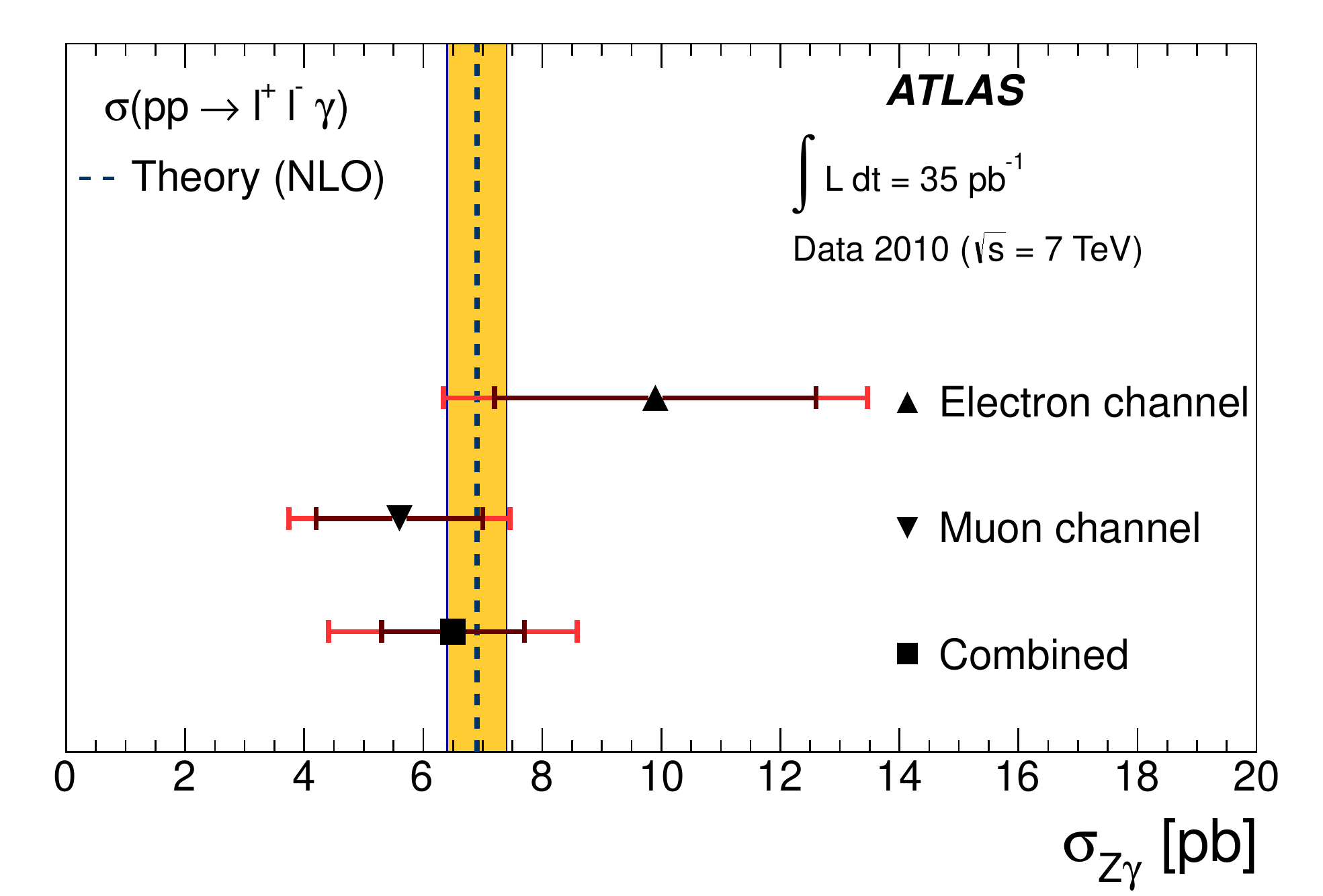}
  \caption{The measured inclusive $W\gamma$ and $Z\gamma$ production cross sections together with SM prediction. Results are shown for the electron and muon final states as well as for their combination. The inner error bar represents the statistical uncertainties and the outer represents the total uncertainties (statistical, systematic and luminosity). All uncertainties are added in quadrature.
The one standard deviation uncertainty in the SM prediction is represented by the vertical band.
}
  \label{fig:totalcs}
\end{figure}

\begin{figure}
  \centering
\includegraphics[width=0.60\columnwidth]{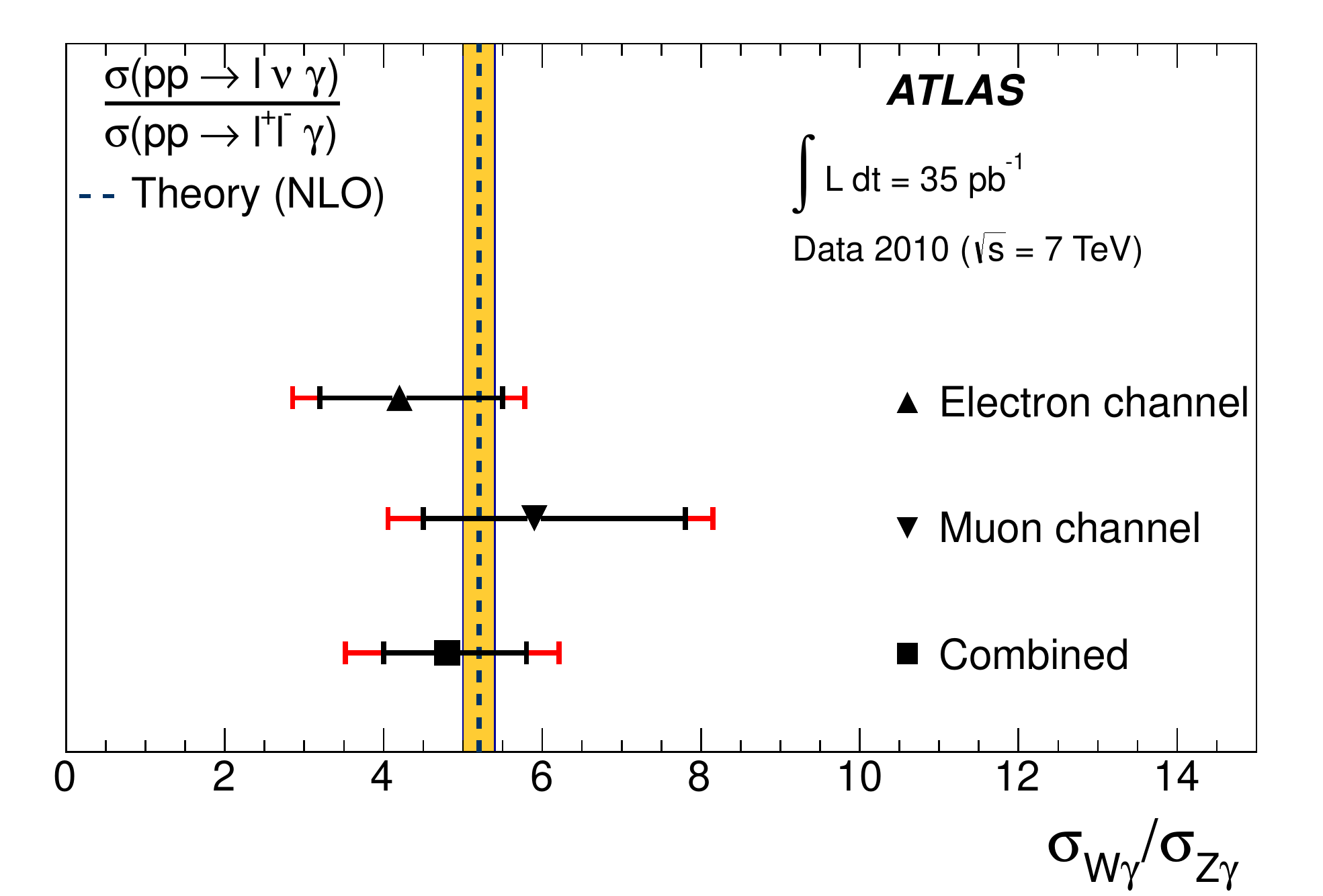}
  \caption{The measured ratio of the production cross sections of $W\gamma$ and $Z\gamma$, together with SM prediction. Results are shown for the electron and muon final states as well as for their combination. The error bars represent the statistical and the total uncertainties. All uncertainties are added in quadrature. The one standard deviation uncertainty in the SM prediction is represented by the vertical band.}
  \label{fig:figratio}
\end{figure}

The Standard Model predictions for the $W\gamma$ and $Z\gamma$
fiducial and production cross sections (as defined in Section
~\ref{sec:fid_cs}) are given in Table~\ref{tab:fidutotal_cs}.
The uncertainty on the cross section predictions includes the following:
\begin{itemize}
\item  The PDF uncertainty is estimated using the MSTW 08 NLO PDF error eigenvectors \cite{pdfmstw} at the 90\% C.L. limit, and variations of $\alpha_s$ in the range from 0.1145 to 0.1176.
\item Renormalisation and factorisation scale uncertainty: this uncertainty is estimated by varying the renormalisation and factorisation scale by factors of two around the nominal scales. 
\item An additional 3\% error is included to account for the approximation of using the $W/Z$ inclusive $k$-factor $k_{\mathrm{FSR}}$ for the $W(Z)\gamma$.

\item Another source of uncertainty accounts for the possible discrepancy
between the photon isolation at the particle level and at the parton level.
Photon isolation at the parton level ($\epsilon_h$), which is implemented in the Baur NLO program as introduced in Section~\ref{sec:dataset}, is used in the calculation of the Standard
Model production cross section predictions. The photon isolation criteria at the particle
level ($\epsilon_h^p$) is used in the acceptance calculation.
 This uncertainty is estimated to be 4\% by
studying the impact on the cross section predicted by the Baur NLO
generator of a 100\% variation of the $\epsilon_h$ parameter.
\end{itemize}

The measured and predicted fiducial and production cross sections of the $pp\to l^{\pm}\nu\gamma+X$ and $pp\to l^+l^-\gamma+X$ processes together with their ratio are shown in Table~\ref{tab:fidutotal_cs} and Table~\ref{tab:tableratio}.

%%%%%%%%%%%%%%%%%%%%%%%%%%%%%%%%%%%%%%%
% TABLE
%%%%%%%%%%%%%%%%%%%%%%%%%%%%%%%%%%%%%%%
\begin{table}
  \centering 
  \begin{tabular}{|c|c|c|}
    \hline
                               &  Experimental measurement &  SM  prediction \\ \hline
                               &  $\sigma^{\mathrm{fid}}[\rm{pb}] $ &  $\sigma^{\mathrm{fid}}[\rm{pb}]$\\ \hline

$pp\rightarrow e^{\pm}\nu\gamma$     & $5.4 \pm 0.7 \pm 0.9 \pm 0.2$   &  $4.7  \pm 0.3 $  \\ 
$pp\rightarrow \mu^{\pm}\nu\gamma$   &  $4.4 \pm 0.6 \pm 0.7 \pm 0.2$  &  $4.9  \pm 0.3 $  \\ 

$pp\rightarrow e^+e^-\gamma$       &  $2.2 \pm 0.6 \pm 0.5 \pm 0.1 $    &  $1.5 \pm 0.1 $  \\ 
$pp\rightarrow \mu^+\mu^-\gamma$   &  $1.4 \pm 0.3 \pm 0.3 \pm 0.1 $    &  $1.7 \pm 0.1 $ \\ \hline

                               &  $\sigma[\rm{pb}] $  &  $\sigma[\rm{pb}]$\\ \hline
$pp\rightarrow e^{\pm}\nu\gamma$     &  $41.1 \pm 5.7 \pm 7.1 \pm 1.4$  & $36.0 \pm 2.3$   \\
$pp\rightarrow \mu^{\pm}\nu\gamma$   &  $33.0 \pm 4.6 \pm 5.5 \pm 1.1$  &  $36.0 \pm 2.3$   \\ 
$pp\rightarrow l^{\pm}\nu\gamma$   &  $36.0 \pm 3.6 \pm 6.2 \pm 1.2$  &  $36.0 \pm 2.3$   \\ 
$pp\rightarrow e^+e^-\gamma$       &   $9.9 \pm 2.7 \pm 2.3 \pm 0.3$    &  $6.9 \pm 0.5$  \\ 
$pp\rightarrow \mu^+\mu^-\gamma$   &  $5.6 \pm 1.4 \pm 1.2 \pm 0.2$    &   $6.9 \pm 0.5$ \\ 
$pp\rightarrow l^+l^-\gamma$   &  $6.5 \pm 1.2 \pm 1.7 \pm 0.2$    &   $6.9 \pm 0.5$ \\ \hline

\hline

  \end{tabular} 

  \caption{Fiducial and production cross sections of the $pp\to l^{\pm}\nu\gamma+X$ and $pp\to ll\gamma+X$ process at $\sqrt{s}=7 \TeV{}$. Both the experimental measurements and the SM NLO predictions are given. The production cross sections are measured with $p_T(\gamma)>15 \GeV{}$, $\Delta R(l,\gamma)>0.7$ and $\epsilon^{p}_{h} < 0.5$, the fiducial cross section is defined in Section~\ref{sec:cs}. For the measurements, the first uncertainty is statistical, the second is systematic and the third is from the luminosity. The uncertainty in the SM prediction is systematic.}
  \label{tab:fidutotal_cs} 
\end{table}

%%%%%%%%%%%%%%%%%%%%%%%%%%%%%%%%%%%%%%%
% TABLE
%%%%%%%%%%%%%%%%%%%%%%%%%%%%%%%%%%%%%%%
\begin{table}
  \centering 
  \begin{tabular}{|c|c|c|}
    \hline
  Cross section                &  Experimental &  SM prediction \\ 
  ratio                         &  measurement &   \\ \hline

  \multicolumn{3}{|c|}{Fiducial phase space }\\
\hline
$\sigma_{pp\rightarrow e^{\pm}\nu\gamma}^{\mathrm{fid}}/\sigma_{pp\rightarrow e^+e^-\gamma}^{\mathrm{fid}}$     & $2.5 _{-0.6}^{+0.8} \pm 0.5 $   &  $3.1  \pm 0.3 $   \\
${\sigma_{pp\rightarrow \mu^{\pm}\nu\gamma}^{\mathrm{fid}}}/{\sigma_{pp\rightarrow \mu^+\mu^-\gamma}^{\mathrm{fid}}}$   &  $3.1 _{-0.8}^{+1.1}  \pm 0.6 $  &  $2.9  \pm 0.3 $  \\ 
\hline
  \multicolumn{3}{|c|}{Phase space for production cross section }\\
\hline
${\sigma_{pp\rightarrow e^{\pm}\nu\gamma}}/{\sigma_{pp\rightarrow e^+e^-\gamma}}$     & $4.2 _{-1.0}^{+1.3}  \pm 0.9 $   &  $5.2  \pm 0.2 $   \\ 
${\sigma_{pp\rightarrow \mu^{\pm}\nu\gamma}}/{\sigma_{pp\rightarrow \mu^+\mu^-\gamma}}$   &  $5.9 _{-1.4}^{+1.9}  \pm 1.2 $  &  $5.2  \pm 0.2 $  \\ 
${\sigma_{pp\rightarrow l^{\pm}\nu\gamma}}/{\sigma_{pp\rightarrow l^+l^-\gamma}}$   &  $4.8 _{-0.8}^{+1.0}  \pm 1.0 $  &  $5.2  \pm 0.2 $  \\ 
\hline
  \end{tabular} 
  \caption{The ratio of $pp\to l^{\pm}\nu\gamma+X$ to $pp\to l^+l^-\gamma+X$ process at $\sqrt{s}=7 \TeV{}$.
  Both the experimental measurement and the SM NLO prediction are given. 
  The production cross sections are measured with $p_\mathrm{T}(\gamma)>15 \GeV{}$, $\Delta R(l,\gamma)>0.7$
  and $\epsilon^{p}_{h} < 0.5$, and the fiducial cross section is defined in Table ~\ref{tab:fiducialcut}.
  The first uncertainty in the experimental measurement is statistical and the second uncertainty is systematic.
  Asymmetric errors calculated from Clopper and Pearson intervals~\cite{stat} are quoted for
  the statistical uncertainty, due to the low statistics in the $pp\to l^+l^-\gamma+X$ measurement.
  The uncertainty in the SM prediction is systematic.}
  \label{tab:tableratio} 
\end{table}

%%%%%%%%%%%%%%%%%%%%%%%%%%%%%%%%%%%%%%%%%%%%%%%%%%%
%% Summary
%%%%%%%%%%%%%%%%%%%%%%%%%%%%%%%%%%%%%%%%%%%%%%%%%%%
%% \input{summary}      % 1 pages

\section{Summary}
\label{sec:summary}

The production processes $pp\rightarrow l^{\pm}\nu\gamma+X$ and
$pp\rightarrow l^{+}l^{-}\gamma+X$ have been studied at $\sqrt{s}=7$ \TeV{} using
$\sim 35$ pb$^{-1}$ of data collected with the ATLAS detector.
The measured fiducial cross sections (defined in the phase-space region where the detector
has good acceptance) and the extrapolated production cross sections
(for $E_{\mathrm{T}}^{\gamma}>15$ \GeV{}, $\Delta R(l,\gamma)>0.7$, and $\epsilon_h^p<0.5$)
for the individual electron, muon and combined decay channels, are presented.
The measurements are in agreement with the predictions of the SM
at $O(\alpha\alpha_s)$ as shown in Table~\ref{tab:fidutotal_cs} and Fig.~\ref{fig:totalcs}.
While the current measurements are not strongly sensitive to possible 
new physics, the distributions of kinematic variables determined from 
the leptons and photons (Figs.~\ref{fig:Wg_kin} and \ref{fig:Zg_kin})
are consistent with the  predictions from the SM in a new kinematic regime,
as is the ratio of the $W\gamma/Z\gamma$ cross sections (Fig.~\ref{fig:figratio}),
which directly depends upon the values of the triple-gauge-couplings in the Standard Model.

%%%%%%%%%%%%%%%%%%%%%%%%%%%%%%%%%%%%%%%%%%%%%%%%%%%
%% acknowledgements
%%%%%%%%%%%%%%%%%%%%%%%%%%%%%%%%%%%%%%%%%%%%%%%%%%%
%% \input{acknowledgments}      % 1 pages

\section*{Acknowledgements}
We gratefully acknowledge the contributions Ulrich Baur made to the
theory calculations used in this study.
We thank CERN for the very successful operation of the LHC, as well as the
support staff from our institutions without whom ATLAS could not be
operated efficiently.

We acknowledge the support of ANPCyT, Argentina; YerPhI, Armenia; ARC,
Australia; BMWF, Austria; ANAS, Azerbaijan; SSTC, Belarus; CNPq and FAPESP,
Brazil; NSERC, NRC and CFI, Canada; CERN; CONICYT, Chile; CAS, MOST and
NSFC, China; COLCIENCIAS, Colombia; MSMT CR, MPO CR and VSC CR, Czech
Republic; DNRF, DNSRC and Lundbeck Foundation, Denmark; ARTEMIS, European
Union; IN2P3-CNRS, CEA-DSM/IRFU, France; GNAS, Georgia; BMBF, DFG, HGF, MPG
and AvH Foundation, Germany; GSRT, Greece; ISF, MINERVA, GIF, DIP and
Benoziyo Center, Israel; INFN, Italy; MEXT and JSPS, Japan; CNRST, Morocco;
FOM and NWO, Netherlands; RCN, Norway; MNiSW, Poland; GRICES and FCT,
Portugal; MERYS (MECTS), Romania; MES of Russia and ROSATOM, Russian
Federation; JINR; MSTD, Serbia; MSSR, Slovakia; ARRS and MVZT, Slovenia;
DST/NRF, South Africa; MICINN, Spain; SRC and Wallenberg Foundation,
Sweden; SER, SNSF and Cantons of Bern and Geneva, Switzerland; NSC, Taiwan;
TAEK, Turkey; STFC, the Royal Society and Leverhulme Trust, United Kingdom;
DOE and NSF, United States of America.

The crucial computing support from all WLCG partners is acknowledged
gratefully, in particular from CERN and the ATLAS Tier-1 facilities at
TRIUMF (Canada), NDGF (Denmark, Norway, Sweden), CC-IN2P3 (France),
KIT/GridKA (Germany), INFN-CNAF (Italy), NL-T1 (Netherlands), PIC (Spain),
ASGC (Taiwan), RAL (UK) and BNL (USA) and in the Tier-2 facilities
worldwide.

%%%%%%%%%%%%%%%%%%%%%%%%%%%%%%%%%%%%%%%%%%%%%%%%%%%
%% Reference
%%%%%%%%%%%%%%%%%%%%%%%%%%%%%%%%%%%%%%%%%%%%%%%%%%%
%%\bibliographystyle{atlasnote}
\bibliographystyle{JHEP}
%% \bibliography{jhep_wzgamma2010}

\providecommand{\href}[2]{#2}\begingroup\raggedright\begin{thebibliography}{10}

\bibitem{DetectorPaper:2008}
{\bf ATLAS} Collaboration, {\it {The ATLAS Experiment at the CERN Large Hadron
  Collider}},  {\em JINST} {\bf 3} (2008) S08003.

\bibitem{CDFpaper}
{\bf The CDF} Collaboration, T.~Aaltonen{~\em et al.}, {\it {Measurement of
  $Z\gamma$ Production in $p\bar{p}$ Collisions at $\sqrt{s}=1.96$~TeV}},
  {\em Phys. Rev.} {\bf D82} (2010) 031103,
  \href{http://xxx.lanl.gov/abs/1004.1140}{{\tt arXiv:1004.1140}}.

\bibitem{D0paper}
{\bf The D0} Collaboration, V.~Abazov{~\em et al.}, {\it First study of the
  radiation-amplitude zero in $w\gamma{}$ production and limits on anomalous
  $ww\gamma{}$ couplings at $\sqrt{s}=1.96$~tev},  {\em Phys. Rev. Lett.} {\bf
  100} (2008) 241805.

\bibitem{Chatrchyan:2011rr}
{\bf CMS} Collaboration, S.~Chatrchyan {\em et.~al.}, {\it {Measurement of
  W-gamma and Z-gamma production in pp collisions at sqrt(s) = 7 TeV}},
  \href{http://xxx.lanl.gov/abs/1105.2758}{{\tt arXiv:1105.2758}}.

\bibitem{madgraph}
J.~Alwall{~\em et al.}, {\it Madgraph/madevent v4: The new web generation},
  {\em JHEP} {\bf 0709} (2007) 028.

\bibitem{pythia}
T.~Sj\"ostrand, S.~Mrenna, and P.~Skands, {\it {PYTHIA} 6.4 physics and
  manual},  \href{http://xxx.lanl.gov/abs/hep-ph/0603175}{{\tt
  hep-ph/0603175}}.

\bibitem{photos}
P.~Golonka and Z.~Was, {\it Photos monte carlo: A precision tool for qed
  corrections in z and w decays},  {\em Eur. Phys. J.} {\bf C45} (2006)
  97--107.

%%%

\bibitem{CTEQ6l1}
J.~Pumplin{~\em et al.}, {\it New generation of parton distributions with
  uncertainties from global qcd analysis},  {\em JHEP} {\bf 0207} (2002) 012.

\bibitem{atlas_tune}
{\bf ATLAS} Collaboration, ``{ATLAS Monte Carlo tunes for MC09}.'' {\scshape
  ATLAS} public note: {\scshape ATL-PHYS-PUB-2010-002},
  http://cdsweb.cern.ch/record/1247375.

\bibitem{PhysRevD.48.5140}
U.~Baur, T.~Han, and J.~Ohnemus, {\it Qcd corrections to hadronic $w\gamma{}$
  production with nonstandard $ww\gamma{}$ couplings},  {\em Phys. Rev.} {\bf
  D48} (1993) 5140--5161.

\bibitem{NLOpaper}
J.~Ohnemus, {\it Order-$\alpha_s$ calculations of hadronic $w\gamma$ and
  $z\gamma$ production},  {\em Phy. Rev.} {\bf D47} (1991) 940.

\bibitem{fregphoton}
P.~Aurenche{~\em et al.}, {\em Recent critical study of photon production in
  hadronic collisions\/},  Phys. Rev. {\bf D73} (2006)  094007.

\bibitem{powheg}
S.~Frixione, P.~Nason, and C.~Oleari, {\it Matching nlo qcd computations with
  parton shower simulations: the powheg method},  {\em JHEP} {\bf 11} (2007)
  070.

\bibitem{atlassimu}
{\bf ATLAS} Collaboration, {\it The atlas simulation infrastructure},  {\em
  Eur. Phys. J.} {\bf C70} (2010) 823--874.

\bibitem{atlaslumi}
{\bf ATLAS} Collaboration, {\it {Luminosity Determination in pp Collisions at
  $\sqrt{s}=7$~TeV Using the ATLAS Detector at the LHC}},  {\em Eur. Phys. J.}
  {\bf C71} (2011) 1630. ATLAS Collaboration, {\it Updated Luminosity
  Determination in pp Collisions at $\sqrt{s}=7$~TeV using the ATLAS Detector},
  ATLAS conference note: {\scshape ATLAS-CONF-2011-011},
  http://cdsweb.cern.ch/record/1334563.

\bibitem{atlaslumi2}
{\bf ATLAS} Collaboration, ``{Updated Luminosity Determination in pp Collisions
  at $\sqrt{s}=7$ TeV using the ATLAS Detector}.'' {\scshape ATLAS} conference
  note: {\scshape ATLAS-CONF-2011-011}, http://cdsweb.cern.ch/record/1334563.

\bibitem{WZpaper}
{\bf ATLAS} Collaboration, {\it Measurement of the $w \rightarrow l\nu$ and
  $z/\gamma* \rightarrow ll$ production cross sections in proton-proton
  collisions at $\sqrt{s} = 7$ \tev{} with the atlas detector},  {\em JHEP}
  {\bf 1012} (2010) 060, \href{http://xxx.lanl.gov/abs/1010.2130}{{\tt
  arXiv:1010.2130}}.

\bibitem{photonpaper}
{\bf ATLAS} Collaboration, {\it Measurement of the inclusive isolated prompt
  photon cross section in pp collisions at sqrt(s) = 7 tev with the atlas
  detector},  {\em Phys. Rev.} {\bf D83} (2011) 052005,
  \href{http://xxx.lanl.gov/abs/1012.4389}{{\tt arXiv:1012.4389}}.

\bibitem{ATLAS_Wjet}
{\bf ATLAS} Collaboration, {\it Measurement of the production cross section for
  $w$-bosons in association with jets in $pp$ collisions at $\sqrt{s}=7$ \tev{}
  with the atlas detector},  {\em Phys. Lett. B} {\bf 698} (2011) 325--345,
  \href{http://xxx.lanl.gov/abs/1012.5382}{{\tt arXiv:1012.5382}}.

\bibitem{jetclean}
{\bf ATLAS} Collaboration, ``{Data-quality requirements and event cleaning for
  jets and missing transverse energy re-construction with the ATLAS detector in
  proton-proton collisions at a center-of-mass energy of $\sqrt{s}=7$ TeV}.''
  {\scshape ATLAS} conference note: {\scshape ATLAS-CONF-2010-038},
  http://cdsweb.cern.ch/record/1277678.

\bibitem{ALPGEN}
M.~L. Mangano{~\em et al.}, {\it Alpgen, a generator for hard multiparton
  processes in hadronic collisions},  {\em JHEP} {\bf 0307} (2003) 001.

\bibitem{atlas_det}
{\bf ATLAS} Collaboration, ``Expected performance of the atlas experiment -
  detector, trigger and physics.'' {\scshape CERN-OPEN-2008-020}.

\bibitem{H1}
{\bf The H1 and ZEUS} Collaboration, F.~D. Aaron{~\em et al.}, {\it Combined
  measurement and qcd analysis of the inclusive $e^{\pm}p$ scattering cross
  sections at hera},  {\em JHEP} {\bf 1001} (2010) 109.

\bibitem{pdfmrst}
A.~Sherstnev and R.~S. Thorne, {\it Parton distributions for lo generators},
  {\em Eur. Phys. J.} {\bf C55} (2008) 553--575.

\bibitem{pdfmstw}
A.~D. Martin, W.~J. Stirling, R.~S. Thorne, and G.~Watt, {\it Parton
  distributions for the lhc},
  \href{http://xxx.lanl.gov/abs/arXiv/0901.0002}{{\tt arXiv/0901.0002}}.

\bibitem{stat}
R.~M. Price and D.~G. Bonett, {\it Estimating the ratio of two poisson rates},
  {\em Computational Statistics \& Data Analysis} {\bf 34} (September, 2000)
  345--356.


\end{thebibliography}\endgroup

\providecommand{\href}[2]{#2}\begingroup\raggedright\endgroup

\clearpage
%\newpage

%%%%%%%%%%%%%%%%%%%%%%%%%%%%%%%%%%%%%%%%%%%%%%%%%%%
%% ATLAS AUTHORS
%%%%%%%%%%%%%%%%%%%%%%%%%%%%%%%%%%%%%%%%%%%%%%%%%%%
%% \input{atlas_authlist}      % 1 pages

% ATLAS Collaboration author list for 12-APR-2011
% Data extracted on 24-May-2011 for paperid 80
%%\documentclass[11pt]{article}
%%\usepackage{a4wide}\begin{document}

\begin{flushleft}
\label{app:collab}
{\Large The ATLAS Collaboration}

\bigskip

G.~Aad$^{\rm 48}$,
B.~Abbott$^{\rm 111}$,
J.~Abdallah$^{\rm 11}$,
A.A.~Abdelalim$^{\rm 49}$,
A.~Abdesselam$^{\rm 118}$,
O.~Abdinov$^{\rm 10}$,
B.~Abi$^{\rm 112}$,
M.~Abolins$^{\rm 88}$,
H.~Abramowicz$^{\rm 153}$,
H.~Abreu$^{\rm 115}$
% ...
\bigskip

% ...
$^{10}$ Institute of Physics, Azerbaijan Academy of Sciences, Baku, Azerbaijan\\
$^{11}$ Institut de F\'isica d'Altes Energies and Universitat Aut\`onoma  de Barcelona and ICREA, Barcelona, Spain\\
% ...
$^{48}$ Fakult\"{a}t f\"{u}r Mathematik und Physik, Albert-Ludwigs-Universit\"{a}t, Freiburg i.Br., Germany\\
$^{49}$ Section de Physique, Universit\'e de Gen\`eve, Geneva, Switzerland\\
% ...
$^{88}$ Department of Physics and Astronomy, Michigan State University, East Lansing MI, United States of America\\
% ...
$^{111}$ Homer L. Dodge Department of Physics and Astronomy, University of Oklahoma, Norman OK, United States of America\\
$^{112}$ Department of Physics, Oklahoma State University, Stillwater OK, United States of America\\
% ...
$^{115}$ LAL, Univ. Paris-Sud and CNRS/IN2P3, Orsay, France\\
% ...
$^{118}$ Department of Physics, Oxford University, Oxford, United Kingdom\\
% ...
$^{153}$ Raymond and Beverly Sackler School of Physics and Astronomy, Tel Aviv University, Tel Aviv, Israel\\
% ...
\end{flushleft}

%%%\end{document}

\end{document}